\documentclass[prd,showpacs,preprint,preprintnumbers,amsmath,amssymb,eqsecnum,nofootinbib]{revtex4-1}
\usepackage{graphicx,amssymb,amsmath,amsthm,amsfonts,mathrsfs,epsfig,epsf}
\usepackage{dcolumn}
\usepackage{amsfonts,mathrsfs}
\usepackage{bm}
\usepackage{subfigure}
\usepackage{color}
\usepackage{url}
\usepackage[dvipdfmx]{hyperref}

\usepackage[top=20truemm,bottom=20truemm,left=20truemm,right=20truemm]{geometry}

\newcommand{\D}{{\rm d}}
\newcommand{\eq}[1]{\mbox{Eq.~(\ref{#1})}}
\newcommand{\fig}[1]{\mbox{Fig.~\ref{#1}}}
\begin{document}

\preprint{RUP-19-14}

\title{Burst particle creation in gravitational collapse to a 
horizonless compact object}

\author{Takafumi Kokubu} 
\email{14ra002a@rikkyo.ac.jp}
\author{Tomohiro Harada}
\email{harada@rikkyo.ac.jp}
\affiliation{Department of Physics, Rikkyo University, Toshima,
Tokyo 171-8501, Japan}
%
%
\date{\today}
%
\begin{abstract}
In the previous paper [Harada, Cardoso, and Miyata, Phys.\ Rev.\ D {\bf
 99} (2019),  044039], it is shown that a hollow transmissive shell
 collapsing to an ultracompact
 object of radius very close to its horizon radius 
generally emits transient Hawking radiation (THR) followed by a couple
 of bursts separated each other by a long time interval.
In the current paper, we expand the previous work in
two independent directions: changing boundary conditions and 
specifying the equations of state (EOSs) of the matter.
First, we introduce a 
 perfectly reflective surface collapsing to an ultracompact object
and find that 
this model also emits THR that is followed only by 
a single burst. Second, we introduce two different collapse dynamics
to an ultracompact object and specify the
corresponding matter EOSs.
We find that THR is quite commonly seen in early times, while 
the subsequent bursts strongly depend on the boundary condition
 and the EOS or the braking behavior of the surface.
\end{abstract}


\maketitle

\tableofcontents

\newpage

\section{Introduction}

In the current 
development of observational and theoretical astronomy and
astrophysics, there is no doubt that black holes play a central
role. The LIGO and Virgo terrestrial interferometric gravitational wave
detectors have directly observed gravitational waves from binary black
holes of tens of solar masses and confirmed the correctness of general
relativity with high accuracy in such a strong field of gravity~\cite{Abbott:2016blz,LIGOScientific:2018jsj}.
More recently, the Event Horizon Telescope has succeeded in imaging the
``shadow'' of a central massive object of M87 as a powerful evidence for
the existence of a supermassive black hole there~\cite{Akiyama:2019cqa}.
On the other hand, these tremendous successes 
of astronomical observation turn the attention of researchers 
to a profound question on the nature of the black hole.
What is the most decisive character 
and signature to distinguish a black hole from 
possible horizonless dark compact massive objects
of radius $R$ slightly larger than the horizon radius, 
$2GM/c^{2}$, where $M$ is the mass of the object?
See Cardoso and Pani~\cite{Cardoso:2019rvt}
for a recent review on horizonless dark compact 
objects and references therein. See 
also~\cite{Mazur:2004fk,Visser:2003ge,Barcelo:2007yk,Visser:2009pw}.
A black hole is defined by an event
horizon, which is the intersection of the past boundary of 
the future null infinity with the spacetime itself. By this definition,
the black hole can never be observed by a distant observer or an observer 
outside the black hole. This means that even if an observer infers
the existence of something very massive in a highly compact region
without any signal of emission from itself, 
she or he can never conclude within a finite time that there is 
an event horizon or a black hole, as is clearly 
demonstrated in~\cite{Nakao:2018knn}.
What the observer outside the horizon can do is to accumulate null
results for the existence of a horizonless compact object 
as long as possible.
This is the strongest evidence of a black hole that the observer
can ever obtain in principle in
classical physics.

In quantum field theory, however, the situation can be very
different because 
the Hawking radiation~\cite{Hawking:1974rv,Hawking:1974sw} can be
regarded as the positive signature of a black hole. 
One might think that the existence of
a black hole can be directly tested by the detection of the Hawking
radiation, which is nearly black-body radiation of all possible degrees
of freedom with temperature $T=T_{H}:=c^{3}\hbar/(8\pi kGM)$. In the previous 
work~\cite{Harada:2018zfg} (hereafter Paper I),
the authors have addressed this
fundamental issue. Using a hollow
transmissive spherical shell, they showed that a collapsing body 
can emit temporarily thermal radiation with $T=T_{H}$ during a
finite time as if it is Hawking radiation, 
even though the collapse does not lead to the formation of 
any kind of horizon but to the formation of an ultracompact
horizonless object. 
They call this radiation transient Hawking radiation (THR). 
The time evolution of radiation deviates from the standard Hawking
radiation when the collapse is significantly slowed down.
The typical feature of the radiation history after the THR 
is a couple of bursts, the first immediately after the THR 
and the second very long after the first.
We should note that the collapse model of a hollow transmissive spherical
shell provides an unique probe into quantum field theory in
gravitational collapse spacetimes as demonstrated  
in Refs.~\cite{Paranjape:2009ib,Banerjee:2009nn,Akhmedov:2015xwa}.

In the current paper, we expand the analysis in Paper
I~\cite{Harada:2018zfg} in two
independent directions. The first direction is to study a different
boundary condition. Here, besides the transmissive boundary
condition, we introduce a perfectly reflective 
boundary condition at the surface of the collapsing body.
In this case, we do not have to specify the interior of the surface
because the quantum field does not propagate in the interior region.
Hence, we can think that it is filled with matter as a usual
astrophysical star. We might heuristically regard
the transmissive boundary condition as that for 
gravitational waves, while the reflective boundary condition as that 
for electromagnetic waves against a conductive surface.
The second direction is to think the matter of
the hollow shell. Because such a static ultracompact object needs to be 
supported against highly relativistic strong gravity, 
the matter field must have very strong pressure.
We address what kind of equation of state (EOS) can explain the assumed 
collapse dynamics to an ultracompact object.

The organization of this paper is as follows.
In Sec.~\ref{sec:particle_creation}, we present the result of 
the application of quantum field theory in 
curved spacetime to an asymptotically flat spherically symmetric spacetime
for both the transmissive and reflective boundary conditions.
In Sec.~\ref{sec:standard_collapse}, we show that the THR is a common
feature of the standard collapse with the 
radius very close to the horizon radius
even if the collapse does not lead to the formation of any horizon.
In Sec.~\ref{sec:qualitative}, we introduce a collapse dynamics model to an ultracompact
object and discuss that a single burst will
be emitted after the THR in the case of the 
reflective boundary condition in contrast to a couple of bursts
separated each other by a very long lapse of time in the transmissive case.
In Sec.~\ref{sec:concrete_models}, we present concrete dynamical
models for the collapse to an ultracompact object in terms of the
effective potential and the EOS of the matter on the shell.
In Sec.~\ref{sec:numerical}, we present numerical results for particle creation 
in the concrete collapse models with both the transmissive and
reflective boundary conditions 
and discuss that the numerical results can be well
explained in terms of the qualitative analysis in Paper I~\cite{Harada:2018zfg}
and in Sec.~\ref{sec:qualitative} in the current paper.
Section~\ref{sec:conclusion} is devoted to summary and conclusion.
We use the units in which $c=G=1$ throughout the paper. 
\section{Particle creation from a spherically symmetric star}
\label{sec:particle_creation}

We calculate quantum particle creation in spherically symmetric
asymptotically flat spacetimes. To do this, we adopt standard 
assumptions and approximations, including geometrical optics
approximation, which are described in Paper I~\cite{Harada:2018zfg}.
We just cite the result for the use in the current paper. 
The particle creation can be calculated by 
a mapping function $G$ defined as $v_{\rm in}=G(u_{\rm out})$,
where $u=u_{out}$ and $v=v_{in}$ are the observer's outgoing null 
coordinate and its corresponding ingoing null coordinate, respectively, 
in terms of the standard double null coordinates $u$ and $v$ in the
asymptotic region. The function $G$ depends on the boundary condition as
we will see below.
The total power of radiation is given by~\cite{Ford:1978ip,Birrell:1982ix}
\begin{equation}
 P\simeq \frac{1}{48\pi}(\kappa^{2}+2\delta \kappa')\, , \label{eq:P_kappa}
\end{equation}
with $\delta=1$ and $0$ for minimally and conformally coupled massless
scalars, respectively, 
where $':=\D/\D u$ and $\kappa(u)$ is defined
as~\cite{Barcelo:2010xk,Kinoshita:2011qs}
\begin{equation}
\kappa(u):=-(\ln G')'.
\end{equation}
We will omit the second term in the parentheses in Eq.~(\ref{eq:P_kappa})
because it does not contribute to the integrated energy of radiation.
This omission leads to the calculation of 
the power for the conformally coupled massless
scalar field.
If the function $\kappa(u_{*})$ 
is positive and satisfies the adiabatic condition
$
|\kappa'(u_{*})|\ll \kappa^{2}(u_{*})\,,
$
then the spectrum of outgoing particles at $u=u_{*}$ can be 
regarded as thermal with temperature 
$T(u_{*})$ such that~\cite{Barcelo:2010xk,Kinoshita:2011qs} 
\begin{equation}
kT(u_{*})=\frac{\kappa(u_{*})}{2\pi}\,.
\label{eq:time-dependent_temperature}
\end{equation}

The function $\kappa(u)$ depends not only on the dynamics of the
spacetime but also the boundary conditions of the stellar surface. 
In Paper I~\cite{Harada:2018zfg}, the authors consider a situation where a hollow spherical shell which is 
perfectly transmissive to scalar waves collapses in vacuum.
In this case, $G'(u)$ and 
$\kappa(u)$ are given by 
\begin{equation}
 G'(u)=\frac{A(\tau_{out})}{B(\tau_{in})},\quad 
 \kappa(u)=C(\tau_{out})-\frac{A(\tau_{out})}{B(\tau_{in})}D(\tau_{in}),
\label{eq:kappa_transmissive}
\end{equation}
where $A(\tau)$, $B(\tau)$, $C(\tau)$, and $D(\tau)$ are given in terms of
$R(\tau)$, $\dot{R}(\tau)$, and $\ddot{R}(\tau)$, where $r=R(\tau)$
 is the radius of the shell as a function of the proper time $\tau$ of
 an observer staying at the surface, 
the dot denotes the derivative with respect to $\tau$,
and $\tau_{out}$ and $\tau_{in}$ are the values of $\tau$ at which 
the outgoing null ray $u=u_{out}$ and the ingoing null ray 
$v=v_{in}$ cross the surface, respectively, as shown in 
Fig.~\ref{fg:v_G_u} (a) or (b).
The explicit expressions for these functions are given in Appendix A of
Paper I~\cite{Harada:2018zfg}. 
\begin{figure}[htbp]
 \begin{center}
\begin{tabular}{cc}
\subfigure[]{\includegraphics[width=0.22\textwidth]{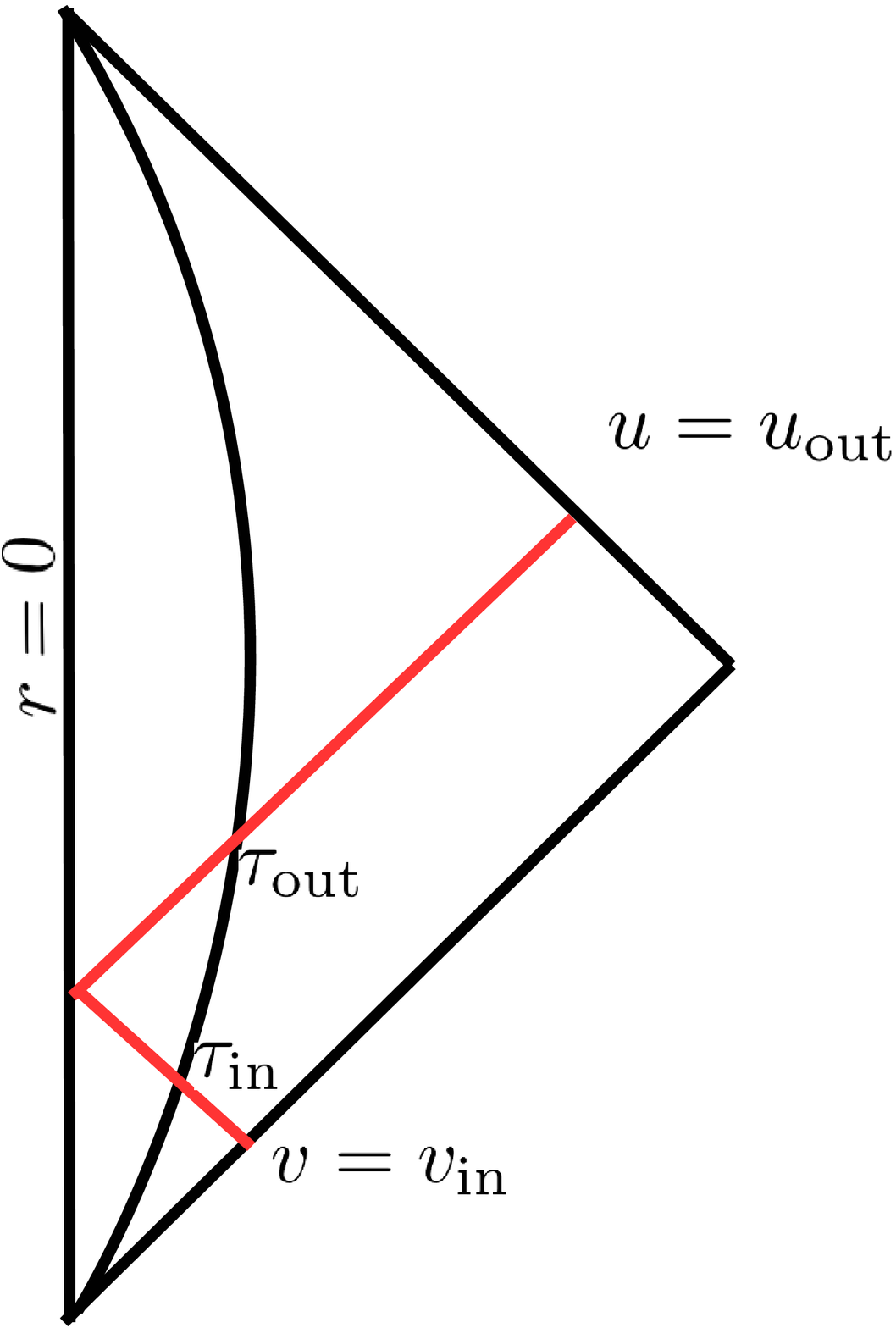}}
 & 
\subfigure[]{\includegraphics[width=0.22\textwidth]{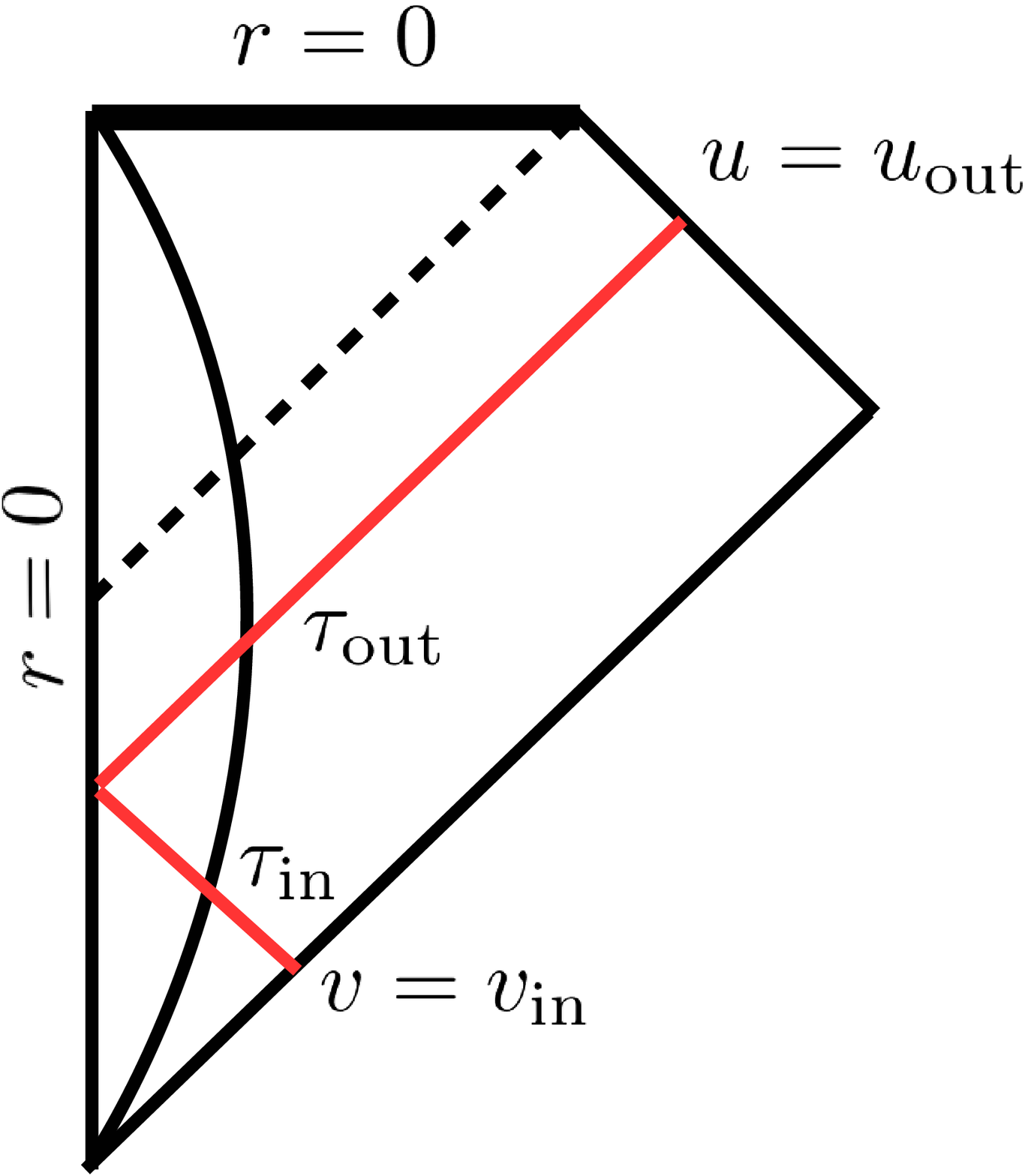}}\\
\subfigure[]{\includegraphics[width=0.22\textwidth]{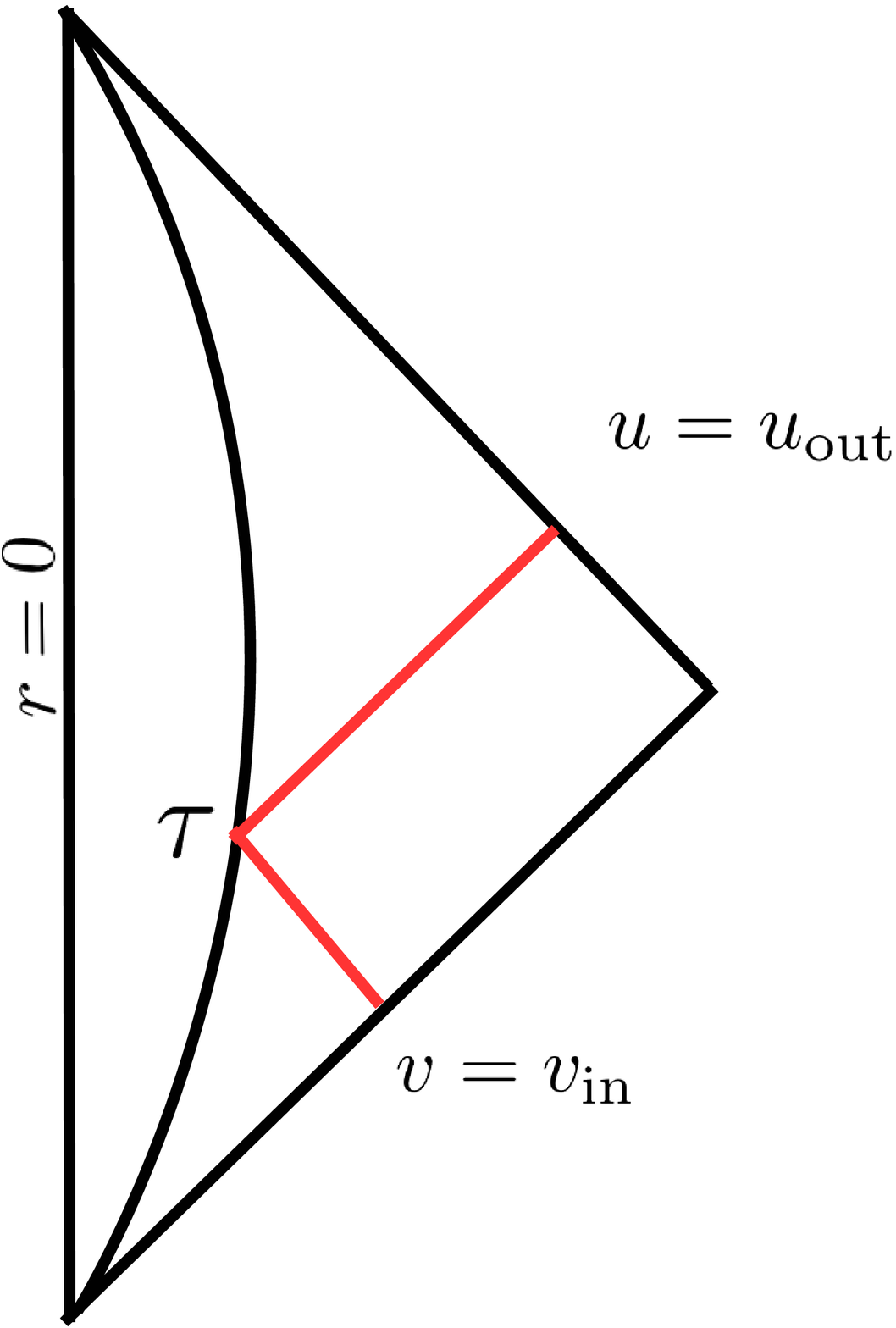}}
 & 
\subfigure[]{\includegraphics[width=0.22\textwidth]{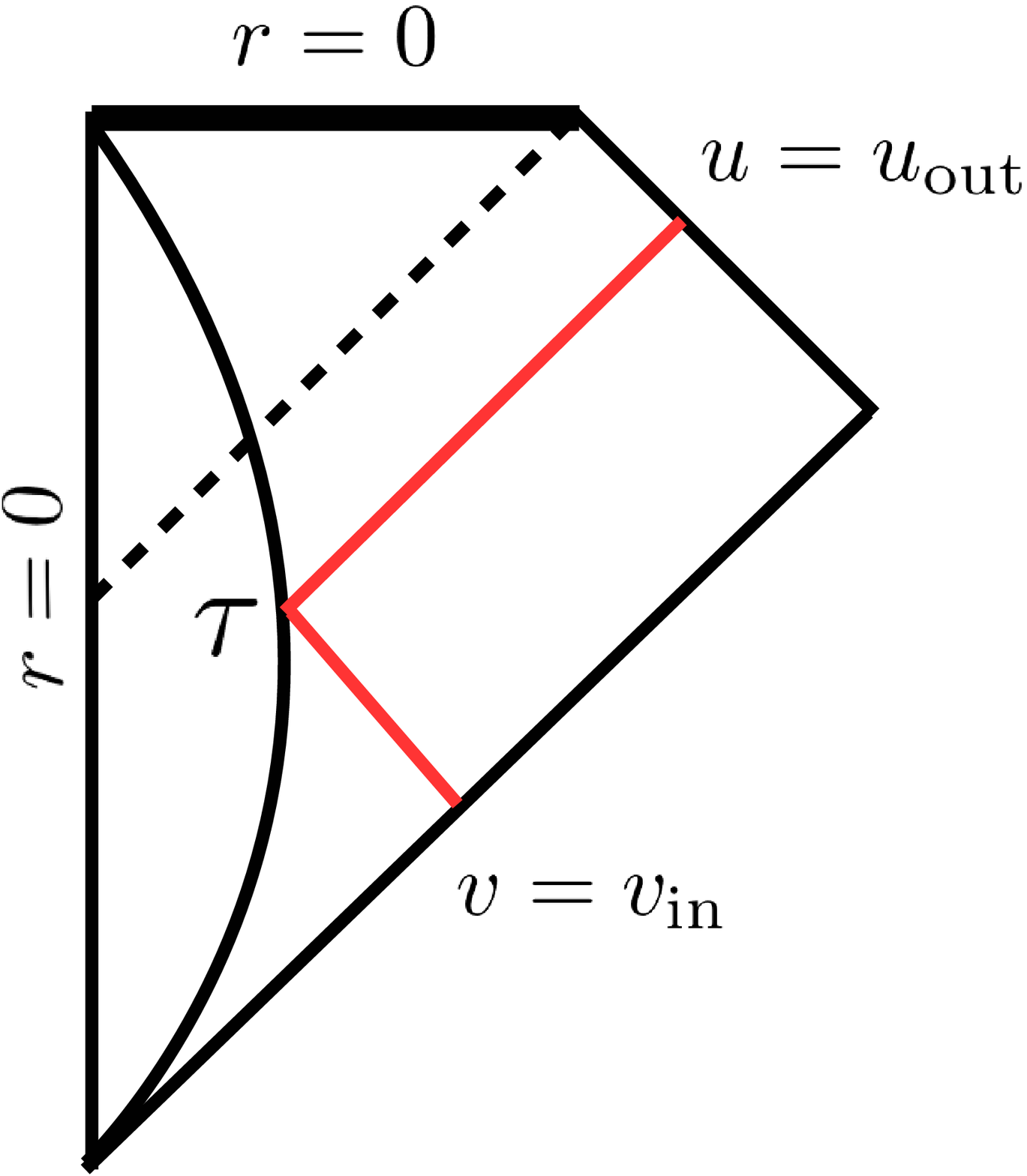}}
\end{tabular}
\caption{
The conformal diagrams for the spacetimes of collapse to a static star
  [(a) and (c)] and a black hole [(b) and (d)]. 
A pair of outgoing and ingoing null
  rays with $u=u_{\rm out}$ and $v=v_{\rm in}$
is depicted for a transmissive surface in (a) and (b),
and for a reflective surface in (c) and (d). 
\label{fg:v_G_u}
}
\end{center}
\end{figure}

Next, we consider a situation where the stellar surface is perfectly reflective against scalar waves. We assume that the exterior is described by the Schwarzschild metric
\begin{equation}
 \D s^{2}=-\left(1-\frac{2M}{r}\right)\D t^{2}+\left(1-\frac{2M}{r}\right)^{-1}\D r^{2}
+r^{2}(\D \theta^{2}+\sin^{2}\theta \D \phi^{2}),
\end{equation}
while the metric in the interior of the star is left arbitrary.
Defining 
\begin{equation}
 u=t-r^{*},\quad v=t+r^{*},
\end{equation}
where $dr^{*}=dr/(1-2M/r)$, 
we find 
\begin{equation}
 \dot{u}=\frac{\sqrt{1-\frac{2M}{R}+\dot{R}^{2}}-\dot {R}}{1-\frac{2M}{R}},\quad 
 \dot{v}=\frac{\sqrt{1-\frac{2M}{R}+\dot{R}^{2}}+\dot {R}}{1-\frac{2M}{R}}\,,\label{eq:u_dot_v_dot}
\end{equation}
from the junction condition of the first fundamental form, 
where $\tau$ is the value of $\tau$ 
at which the ingoing null ray $v=v_{in}$ is reflected by the
surface to the outgoing null ray $u=u_{out}$
as seen in Fig.~\ref{fg:v_G_u} (c) or (d). 
We do not need two proper times 
$\tau_{out}$ and
$\tau_{in}$ but only $\tau$ in the reflective case. Then, we find 
\begin{equation}
 G'(u_{out})=\frac{\dot{v}}{\dot{u}}(\tau),~~
 \kappa(u)=\frac{1}{\dot{u}}\frac{d}{d\tau}\ln\frac{\dot{u}}{\dot{v}}(\tau).
\end{equation}
Implementing the calculation, we find
\begin{eqnarray}
G'= \frac{\sqrt{1-\frac{2M}{R}+\dot{R}^{2}}+\dot{R}}{\sqrt{1-\frac{2M}{R}+\dot{R}^{2}}-\dot{R}},~~
 \kappa=
2\frac{\sqrt{1-\frac{2M}{R}+\dot{R}^{2}}+\dot{R}}{\sqrt{1-\frac{2M}{R}+\dot{R}^{2}}}\left(-\ddot{R}+\frac{M}{R^{2}}\frac{\dot{R}^{2}}{1-\frac{2M}{R}}\right),
\label{eq:kappa_reflective}
\end{eqnarray}
where the right-hand sides are evaluated at the retarded time $\tau$ and
we omit it for brevity.

Therefore, only the state of the shell at the retarded time $\tau$ is
responsible for particle creation at $u=u_{out}$ unlike in the transmissive case.
If $M=0$, Eq.~(\ref{eq:kappa_reflective}) reduces to 
\begin{equation}
 \kappa=-2\left(1+\frac{\dot{R}}{\sqrt{1+\dot{R}^{2}}}\right)\ddot{R}.
\end{equation}
This corresponds to particle creation by a spherical moving mirror in the flat
spacetime, which is comparable with a planar moving mirror in the flat
spacetime discussed in~\cite{Birrell:1982ix}.
If the surface is expanding with highly relativistic
speed, i.e., $\dot{R}\gg 1$, then Eq.~(\ref{eq:kappa_reflective})
reduces to 
\begin{equation}
 \kappa\simeq \frac{4M}{R^{2}\left(1-\frac{2M}{R}\right)}\Gamma^{2}-4\ddot{R},
\end{equation}
where $\Gamma:=[1-(dR/dt)^{2}]^{-1/2}=\sqrt{1+\dot{R}^{2}}$. 
In the above, the first term
gives an interesting effect: If $\Gamma\gg 1$, the first term can be
very large and, hence,  
$\kappa$ can be much greater than its Hawking value $\kappa_{H}=1/(4M)$
independently from the acceleration even if $R>2M$.
On the other hand, if the surface is shrinking with highly relativistic
speed, i.e., $-\dot{R}\gg 1$, then Eq.~(\ref{eq:kappa_reflective})
reduces to 
\begin{equation}
 \kappa\simeq -\left(1-\frac{2M}{R}\right)\frac{\ddot{R}}{\Gamma^{2}}+\frac{M}{R^{2}}.
\end{equation}
Therefore, there is not enhancement but suppression on the first term 
due to a relativistic speed.

\section{Transient Hawking radiation}
\label{sec:standard_collapse}

In Paper I~\cite{Harada:2018zfg}, particle creation is calculated 
for a hollow 
transmissive shell. Here, we calculate particle creation for a
perfectly reflective surface.
First, we consider particle creation from standard collapse.

By the standard collapse, we refer to the following dynamics:
\begin{enumerate}
 \item Phase 0, an early-collapse phase: $\tau<\tau_{0}$ or $u<u_{0}$.\\
We assume
\begin{equation}
1-\frac{2M}{R}> \frac{1}{2}, \quad |\dot{R}|\lesssim 1, \quad
 \mbox{and}\quad 
 |\ddot{R}|\lesssim \frac{1}{2M}.
\end{equation}
 \item Phase 1, a late-collapse phase: $\tau>\tau_{0}$ or $u>u_{0}$.\\
We assume
\begin{equation}
 1-\frac{2M}{R}< \frac{1}{2},~~ 1-\frac{2M}{R}< \dot{R}^{2},~~\dot{R}=O(1),~~ \mbox{and}~~ \ddot{R}=O((2M)^{-1}).
\end{equation}
\end{enumerate}

From Eq.~(\ref{eq:kappa_reflective}), we can estimate $\kappa(u)$ as follows:
\begin{enumerate}
\item $u<u_{0}$\\
We have 
\begin{eqnarray}
\kappa(u)\simeq   2\left[M\left(\frac{\dot{R}}{R}\right)^{2}-\ddot{R}\right].
\end{eqnarray}
Therefore, we can conclude that $|\kappa|\lesssim 1/(4M)$.
Thus, the radiation for $u<u_{0}$, which may be called pre-Hawking radiation, 
is not much stronger than the standard Hawking radiation. 

 \item $u>u_{0}$\\
We have
\begin{equation}
\kappa(u)\simeq \frac{1}{4M}\,.\label{eq:kappa_standard_collapse}
\end{equation}
Since we can also see that the adiabatic condition is satisfied,
the radiation is temporarily thermal with $kT=1/(8\pi M)$.
Therefore, there is THR also in the 
case of a reflective surface. 
Since THR is derived without assuming any horizon, 
we can conclude that THR does not need any horizon.
If the late-collapse phase continues up until $R\simeq
2M(1+\epsilon^{2})$, then the THR arises and
lasts for $\Delta u \simeq 4M \ln \epsilon^{-2}$.
In the limit $\epsilon\to 0$,  
the Hawking radiation continues eternally
and the energy radiated goes to infinity.
\end{enumerate}
These features are common to both transmissive 
and reflective cases.

\section{Particle creation: a qualitative analysis}
\label{sec:qualitative}
In~Ref.\cite{Paranjape:2009ib} and Paper I~\cite{Harada:2018zfg}, 
the authors consider a collapse model in which a static surface 
turns to an imploding null surface at $u=u_{0}$ and then to a static surface
again at $u=u_{1}>u_{0}$ under the transmissive condition, for which 
they find that the THR is derived fully analytically.
In the reflective case, however, this model does not work.
This is because in the imploding phase, $v_{in}=$const. for 
$u_{0}<u_{out}<u_{1}$ and, hence, $G'=0$
for this range of $u_{out}$. Thus, we cannot obtain any meaningful 
function for $\kappa=-(\ln G')'$.
For this reason, we concentrate on the timelike surface model. 
Since the transmissive condition is studied in detail in Paper I~\cite{Harada:2018zfg}, 
we concentrate on the reflective case.

\subsection{Phases of collapse dynamics}

As in Paper I~\cite{Harada:2018zfg}, we assume the following collapse dynamics, after which
a static ultracompact object of radius $R=R_{f}=2M(1+\epsilon^{2})$ is formed.
\begin{enumerate}
\item Phase 0, an early-collapse phase: $\tau<\tau_{0}$ or $R>4M$.\\ 
We assume $1-2M/R> 1/2$, $|\dot{R}|\lesssim 1$, and $|\ddot{R}|\lesssim 1/(2M)$.

\item Phase 1, a late-collapse phase: $\tau_{0}<\tau<\tau_{1}$ or $R_{b}<R<4M$.\\
We assume $1-2M/R< 1/2$, $1-2M/R< \dot{R}^{2}$, $\dot{R}=O(1)$, and
      $\ddot{R}=O((2M)^{-1})$. We denote the end of the standard
      collapse with $\tau_{1}$ and $R_{b}:=R(\tau_{1})$.

\item Phase 2, an early-braking phase: $\tau_{1}<\tau<\tau_{2}$ or $R_2<R<R_b$.\\ 
We assume that at $\tau=\tau_{1}$ or $R=R_b$, the surface begins to brake.
For $\tau_{1}<\tau<\tau_{2}$, we assume the following inequality:
\begin{eqnarray}
1-\frac{2M}{R}< \dot{R}^{2}\, .
\label{eq:hierarchy}
\end{eqnarray}

\item Phase 3, a late-braking phase: $\tau_{2}<\tau<\tau_{3}$ or $R_f< R<R_2$\\
We assume that at $\tau=\tau_{2}$, when $R=R_{2}$, the following
      equality holds: 
\begin{eqnarray}
1-\frac{2M}{R}= \dot{R}^{2}\, .
\label{eq:hierarchy_marginal}
\end{eqnarray}
For $\tau_{2}<\tau<\tau_{3}$, the following inequality holds: 
\begin{eqnarray}
1-\frac{2M}{R}> \dot{R}^{2}\, .
\label{eq:hierarchy_opposite}
\end{eqnarray}
The radius approaches the final value $R_{f}$.

\item Phase 4, a final static phase: $\tau>\tau_{3}$ or $R=R_f$.\\
We assume that $R(\tau)$ smoothly reaches $R_{f}$ at $\tau=\tau_{3}$.
Later on, the surface is completely static. 
\end{enumerate}

\begin{figure}[htbp]
 \begin{center}
  \includegraphics[width=0.3\textwidth]{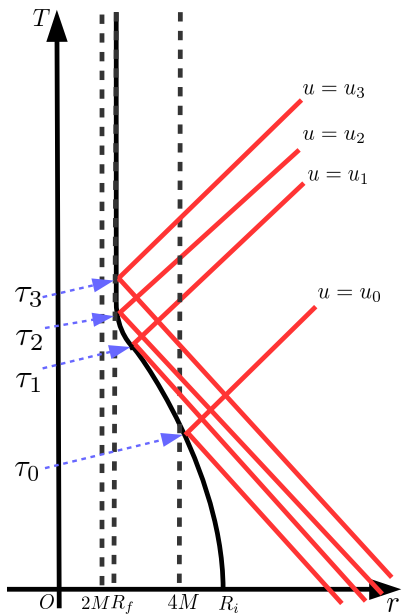}
\caption{
\label{fg:timelike_shell_model_ref}
The collapse model with a timelike surface. The surface enters $R=4M$ at
  $\tau=\tau_{0}$, begins to brake at $\tau=\tau_{1}$, and stops at
  $\tau=\tau_{3}$. Between $\tau_{1}$ and $\tau_{3}$, there is a moment
  $\tau_{2}$, when the equality $1-2M/R=\dot{R}^{2}$ is satisfied. 
The outgoing null rays which
leave the surface outwardly to the Schwarzschild region at $\tau=\tau_{0}$,
  $\tau_{1}$, $\tau_{2}$, and $\tau_{3}$ are denoted by red lines
  labeled $u=u_{0}$, $u_{1}$, $u_{2}$, and $u_{3}$, 
respectively. 
}
 \end{center}
\end{figure}

Figure~\ref{fg:timelike_shell_model_ref} depicts the situation, where 
we label as $u=u_{0}$,
$u_{1}$, $u_{2}$, and $u_{3}$
those outgoing null rays in the Schwarzschild region which leave the surface
outwardly at $\tau=\tau_{0},\,\tau_{1},\, \tau_{2}$, and
$\tau_{3}$, respectively.
 
\subsection{Post-Hawking burst \label{subsec:post-Hawking_burst}}

In Paper I~\cite{Harada:2018zfg}, the authors show that in the transmissive case, 
there are a couple of bursts, 
the first one immediately after the THR and the second one long after
the first. 
Here, we find that in the reflective case, 
there can appear only a single burst of radiation, which is 
immediately after the THR, in contrast to the transmissive case.

For $u_{1}<u<u_{3}$, the observer receives an 
outgoing null ray which left the surface in the braking
phase and can be traced back to the ingoing null ray which reaches
the surface at the same time. The function 
$\kappa(u)$ is approximately given by
\begin{eqnarray}
 \kappa(u)&\simeq &
  -\frac{1-\frac{2M}{R}}{\dot{R}^{2}}\ddot{R}+\frac{1}{4M}, \label{eq:kappa_ref_u1_u2}\\
 \kappa(u)&\simeq &
  -2\ddot{R}+\frac{2\dot{R}^{2}}{1-\frac{2M}{R}}\frac{1}{4M},
\label{eq:kappa_ref_u2_u3}
\end{eqnarray}
for $u_{1}<u<u_{2}$ and $u_{2}<u<u_{3}$, respectively.
Note that the function $\dot{R}^{2}/(1-\frac{2M}{R})$ is generally  
a decreasing function for $u_{1}<u < u_{3}$, which is greater than
unity at $u=u_{1}$, unity at $u=u_{2}$, and zero at $u=u_{3}$.
In Eqs.~(\ref{eq:kappa_ref_u1_u2}) and (\ref{eq:kappa_ref_u2_u3}), 
the second term can be regarded as 
the THR, which keeps constant 
for $u_{1}<u<u_{2}$ and decays for $u_{2}<u<u_{3}$. 
This implies that $u_{2}$ (or $\tau_{2}$) plays a clear physical 
role: it triggers the interruption of THR.
On the other hand, the first term, which is negative, 
 dominates the second term if $\ddot{R}\gtrsim 1/(2M)$  
 at $\tau=\tau_{2}$ or $u=u_{2}$. 
 The emission due to the first term completely ends at $u=u_{3}$.
 This gives a burst of radiation in the end of the
 THR around at
 $u=u_{2}$, which we call a post-Hawking burst. 
 The details of the burst depend
 on the behavior of the surface in the braking phase.
 It is clear that in the reflective case, there is no 
 further burst in contrast to the transmissive case.

\subsection{Time dependence of particle
  creation\label{subsec:specific_models}}
Here we analyze the time dependence of the power of radiation
in detail for a couple of very simple braking models.

\subsubsection{Model A: exponentially slowed-down model}
First, we assume that $R-R_{f}\propto e^{-\sigma\tau}$ for
$\tau_{1}<\tau <\tau_{3}'$ 
by introducing the deceleration parameter $\sigma$ such that
$\ddot{R}= \sigma |\dot{R}|= \sigma^{2}(R-R_{f})$ with 
\begin{equation}
 \sigma= \frac{|\dot{R}_{b}|}{R_{b}-R_{f}},
\label{eq:sigma}
\end{equation}
where $|\dot{R}_{b}|=O(1)$. Then, 
for $\tau'_{3}<\tau<\tau_{3}$, we assume that $R$ smoothly settles 
down to the final fixed radius $R_{f}$ at $\tau=\tau_{3}$. 
We parameterize $\sigma$ such that 
\begin{align}
\sigma^{-1}=2M\epsilon^{2\beta}, \label{sigma-beta}
\end{align} 
where we assume $\beta\ge 1/2$.
The dynamics of this model is described in detail in Paper I~\cite{Harada:2018zfg}.

We can see that the first term for $u_{1}<u<u_{2}$
on the right-hand side of Eq.~(\ref{eq:kappa_ref_u1_u2}) is negative 
and its absolute value increases and reaches 
$O(\epsilon^{1-2\beta}M^{-1})=O(\epsilon\sigma)$
at $u=u_{2}$. The first term on the right-hand side of
Eq.~(\ref{eq:kappa_ref_u2_u3}) then just decays to zero for $u_{2}<u<u_{3}$.
Therefore, the post-Hawking burst is 
much stronger than the THR at $u=u_{2}$ only if $\beta>1/2$, while it is
as strong as the THR if $\beta=1/2$. 
The duration of the burst is given by $u_{3}-u_{1}\simeq
4M(1-\beta)\ln\epsilon^{-2}$ for $1/2\le \beta<1$, $\simeq 4M$ for
$\beta=1$, and $\simeq 4M\epsilon^{2(\beta-1)}$ for $\beta>1$.

\subsubsection{Model B: constant-deceleration model}
Next we consider a technically simpler model, where the deceleration $a$ of the
surface is constant for $\tau_{1}<\tau<\tau_{3}$ with
\begin{equation}
a=\frac{\dot{R}_{b}^{2}}{2(R_{b}-R_{f})},
\label{eq:a}
\end{equation}
where $|\dot{R}_{b}|=O(1)$.
We can naturally assume $a\gg
1/(4M)$. 
We parameterize $a^{-1}=2M\epsilon^{2\beta}$ as in the previous model.
The dynamics of this model is described in detail in Paper I~\cite{Harada:2018zfg}.

We can see that the first term 
on the right-hand side of Eq.~(\ref{eq:kappa_ref_u1_u2}) is negative 
and its absolute value increases for $u_{1}<u<u_{2}$ and reaches 
$O(\epsilon^{-2\beta}M^{-1})=O(a)$
at $u=u_{2}$. The first term on the right-hand side of
Eq.~(\ref{eq:kappa_ref_u2_u3}) is then constant for $u_{2}<u<u_{3}$.
Therefore, the post-Hawking burst is much stronger 
than the THR for $u_{2}<u<u_{3}$ only if $\beta>0$, while it is as
strong as the THR if $\beta=0$.
The duration of the burst is given by $u_{3}-u_{1}\simeq
4M(1-\beta)\ln\epsilon^{-2}$ for $1/2\le \beta<1$, $\simeq 4M$ for
$\beta=1$, and $\simeq 4M\epsilon^{2(\beta-1)}$ for $\beta>1$.

\section{Concrete dynamical models}
\label{sec:concrete_models}

\begin{figure}[t]
\begin{center}
\includegraphics[clip,width=0.5\textwidth]{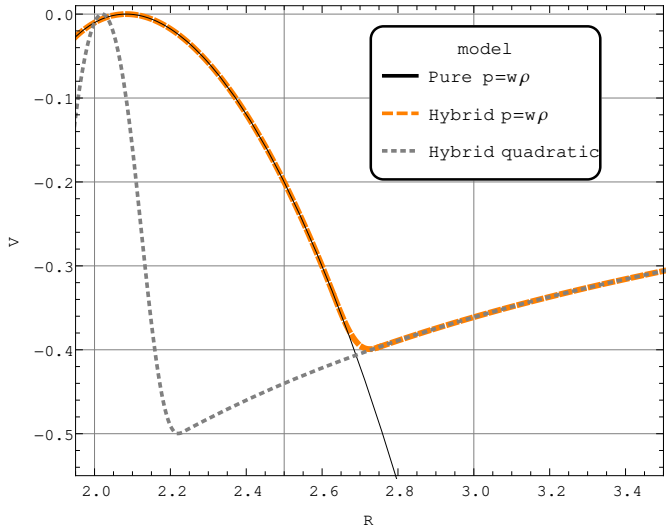}
\caption{
The effective potentials $V(R)$ are plotted with solid, long-dashed, and
 short-dashed lines for 
the pure $p=w\rho$ with $w=1$, hybrid $p=w\rho$ with $w=1$, and hybrid
 quadratic models with $(\epsilon, \beta)=(10^{-1},0.5)$, respectively.
\label{fig:fig-POTENTIALs}}
\end{center}
\end{figure}

In this section, we introduce three 
concrete matter models for the collapsing surface.
All of the models introduced here are categorized as model A in the
previous section, i.e., the surface is 
exponentially slowed down in late times.  
Three effective potentials introduced in this section are summarized in
Fig.~\ref{fig:fig-POTENTIALs}. The details will be discussed 
below for each model.

\subsection{Pure $p=w\rho$ model}\label{section-linear}

When the EOS is $p=w\rho$, where $p$ is the surface
pressure, $\rho$ is the surface energy density, and $w$ is a constant,
the conservation of the kinetic energy of the shell is given by
\begin{align}
\left(\frac{\D  R}{\D \tau}\right)^2+V(R)=0  \quad {\rm with} \quad
V(R)=1-\frac{M}{R} -\left(\frac{m}{2R^{2w+1}}\right)^2 - \left(\frac{MR^{2w}}{m}\right)^2,
\label{re-effective-potential}
\end{align}
where $m$ is a conserved quantity of the shell.
This equation comes from the junction condition of the second
fundamental form and 
the derivation of the potential is delegated to Appendix A.
The equation for extremal values is given by $\D V/\D R=0$.
For $w> 0$, there is a single extremum, for which the 
radius $R=R_{f}$ is given by
\begin{align}
\frac{R_f}{M}=\left[\frac{1+2w}{4w} \left(\frac{m}{M^{1+2w}}\right)^2\right]^{1/(1+4w)} \qquad (w>0). \label{V-primeis0}
\end{align}
Since there is no extremum for $w\le 0$, we concentrate on $w>0$. 
This single extremum gives a global maximum of $V(R)$.
By substituting \eq{V-primeis0} into $V(R_f)$ given by \eq{re-effective-potential}, we obtain $V(R_f)=1-(1+4w)^2(4w)^{-1}(1+2w)^{-1}R_f^{-1}M$.
From the staticity condition $V(R_{f})=0$, 
$R_{f}$ can also be written by
\begin{align}
\frac{R_f}{M}=2+\frac{1}{4w(1+2w)}. \label{Vis0}
\end{align}
The combination of Eqs.~(\ref{V-primeis0}) and (\ref{Vis0}) gives a relation between $ m$ and $w$ as
\begin{align}
\frac{m}{M^{1+2w}}=\frac{(1+4w)^{1+4w}}{(4w)^{2w}(1+2w)^{1+2w}}=:m_f. 
\label{static-condition}
\end{align}
Under the above conditions, there is a unique static solution at $R=R_{f}$.
The time evolution of the shell, which is collapsing at $R>R_{f}$, 
approaches this unstable static radius $R=R_f$.

\begin{figure}[t]
\begin{center}
\includegraphics[clip,width=0.5\textwidth]{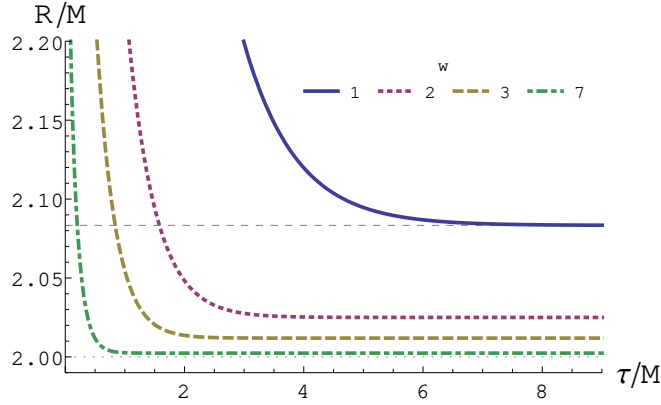}
\caption{The radius $R$ of the shell with the EOS $p=w\rho$ is
 plotted as a function of $\tau$ for different values of $w$ in the
 pure $p=w\rho$ model. 
 The value $w=1$ is the case where the
 dominant energy condition is marginally satisfied. In this case, the
 radius approaches $(25/12) M\simeq 2.08M$, which is plotted with 
 the dashed straight line. For $w>1$, where the dominant energy
 condition is violated, the larger the value of $w$, the closer to $2M$
the final radius.}
\label{fig-timeevolution} 
\end{center}
\end{figure}

Since $R_f$ is a monotonically decreasing
function of $w$ as is seen from \eq{Vis0}, as long as we respect energy
conditions (the null, weak, dominant and strong energy conditions),
i.e., $0<w\leq 1$, $R_f$ has its minimum $(25/12) M \simeq 2.0833M$ at $w=1$.
On the other hand, 
$R_f$ approaches $2M$ as $w\to\infty$. 
Figure \ref{fig-timeevolution} shows the time evolution of the shell
with different values of $w$, 
which is obtained by integrating \eq{re-effective-potential} with \eq{static-condition}.

For $w\gg 1$, $R_f$ and $ m_f$ are approximately given by 
\begin{align}
\frac{R_f}{2M}-1\simeq \left(\frac{1}{4w}\right)^2, \quad
m_f \simeq 2^{1+2w}. \label{large-w-relation}
\end{align}
The behavior of the shell near the final radius $R_f$ is obtained by
using the Taylor expansion of $V(R)$ around $R=R_f$. Since
$V(R_f)=V'(R_f)=0$, \eq{re-effective-potential} reduces to  
$\dot R^2\simeq -V''(R_f)(R-R_f)^2 /2$.
For a contracting shell for $w\gg 1$, this equation can be integrated to give
\begin{align}
R-R_f\simeq De^{-w{\tau}/{M}},
\label{eq:exponential_decay}
\end{align}
where $D$ is a constant of integration. 

\begin{figure}[htbp]
  \begin{center}
    \begin{tabular}{c}

      \begin{minipage}{0.35\hsize}
        \begin{center}
          \includegraphics[width=0.5\textwidth]{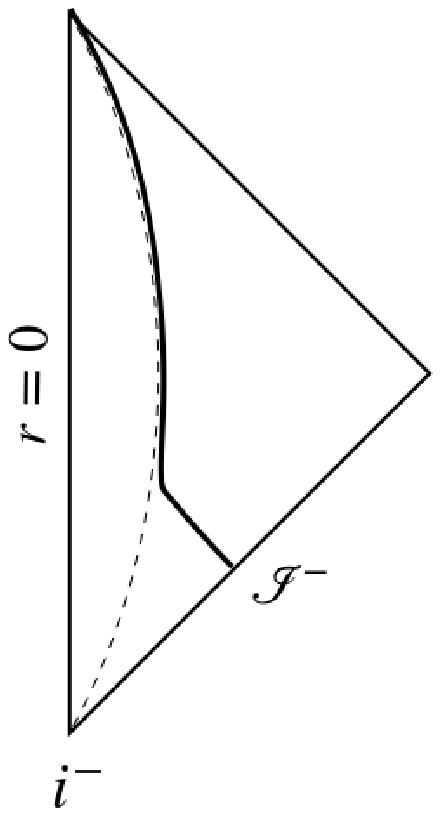}\\ (a)
        \end{center}
      \end{minipage}
 
      \begin{minipage}{0.35\hsize}
        \begin{center}
          \includegraphics[width=0.5\textwidth]{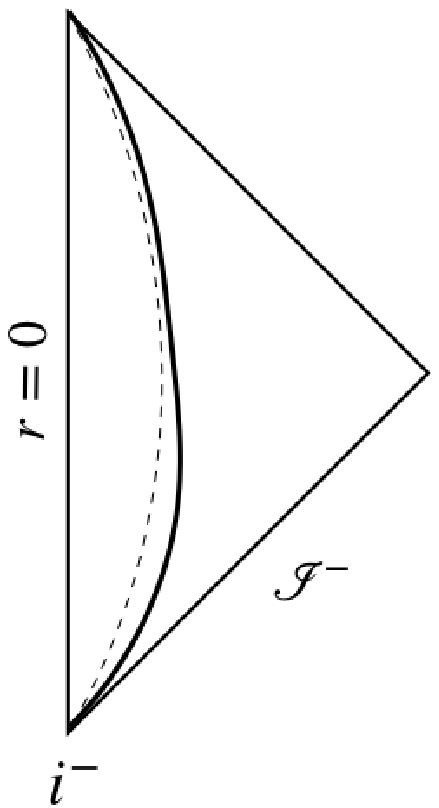}\\(b)
        \end{center}
      \end{minipage}
        
    \end{tabular}
\caption{
(a) The conformal diagram for a collapsing shell in the pure $p=w\rho$
   model. The thick curve depicts the motion of the shell starting from
   the past null infinity $\mathcal I^-$. The shell asymptotically
   approaches the final static radius $R_{f}$, which is slightly larger
   than $2M$. The timelike curve $r=R_{f}$ is depicted by the dotted line.
(b) Same as (a) but for a collapsing shell starting from the past
   timelike infinity $i^-$, which is realized in the hybrid models.
 \label{fig-null-timelike}}
  \end{center}
\end{figure}

We should also note that since $V(R)\to -\infty$ as $R\to \infty$, 
the speed of the shell approaches that of light with respect to
the observer at $r=$const. and, hence, it approaches not the timelike
infinity but the null infinity. 
Thus, the shell emerges from the past null infinity for the dynamics
collapsing from
infinity to $R=R_{f}$ as depicted in \fig{fig-null-timelike} (a).

As defined in Ref.~\cite{Paranjape:2009ib,Banerjee:2009nn} and Paper
I~\cite{Harada:2018zfg}, 
the parameter $\epsilon$ controls the closeness of the final static 
radius to the horizon radius as 
\begin{align}
\frac{R_{f}}{2M}-1=\epsilon^{2}. \label{Rf-epsilon}
\end{align}
Therefore, for $w\gg 1$, Eqs.~(\ref{large-w-relation}),
(\ref{eq:exponential_decay}),
and (\ref{Rf-epsilon}) imply that this model in late times falls into model A
with $\epsilon = 1/(4w)$ and $\sigma=1/(4\epsilon M)$.

\subsection{Hybrid $p=w\rho$ model}\label{section-combined}
As seen in Sec. \ref{section-linear}, a collapsing shell with $p=w\rho$
emerges from the past null infinity. 
However, in astrophysically realistic situations, the star begins to
collapse from a state nearly in free fall or in equilibrium. 
Therefore, we here introduce concrete models where the 
collapsing shell starts from the past timelike infinity. 
The qualitative analysis of this picture without specific model is
studied in detail in Paper I~\cite{Harada:2018zfg} and in
Sec.~\ref{sec:qualitative} in the current paper.
To construct concrete pedagogical models, we look for an EOS
for a collapsing shell which falls from a static state at 
the infinity and approaches the
final radius which is close to its horizon radius.
Since no physical EOS realizing such a dynamics
is known to our knowledge, we reconstruct the EOS from 
the shape of the effective potential for the desired dynamics, 
instead of specifying the explicit form of EOS. 
Once the potential is specified, the corresponding EOS is determined
through 
Eqs.~(\ref{JUNCTION1}) and (\ref{JUNCTION2}).

\begin{figure}
        \begin{center}
\begin{tabular}{cc}
          \subfigure[]{\includegraphics[width=0.45\textwidth]{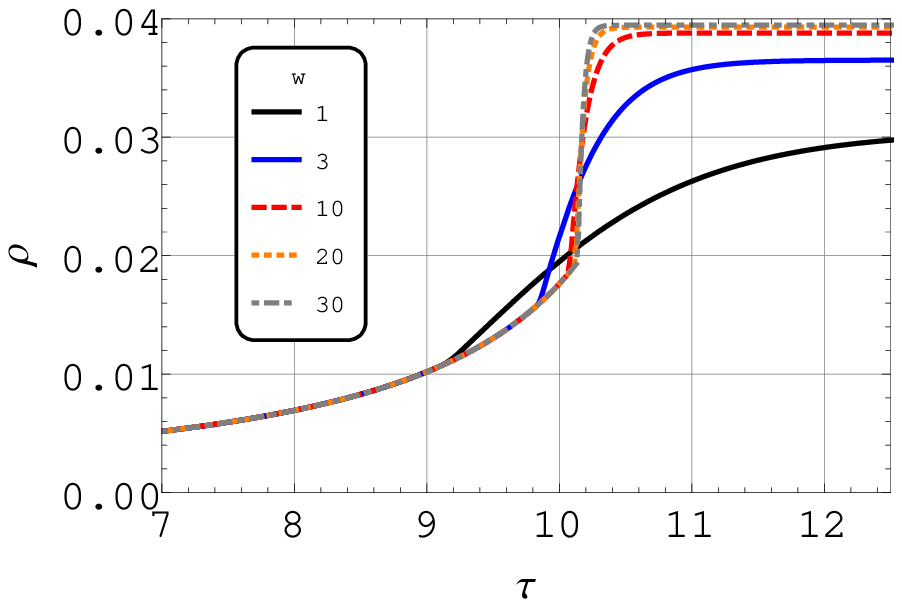}}
 & 
          \subfigure[]{\includegraphics[width=0.45\textwidth]{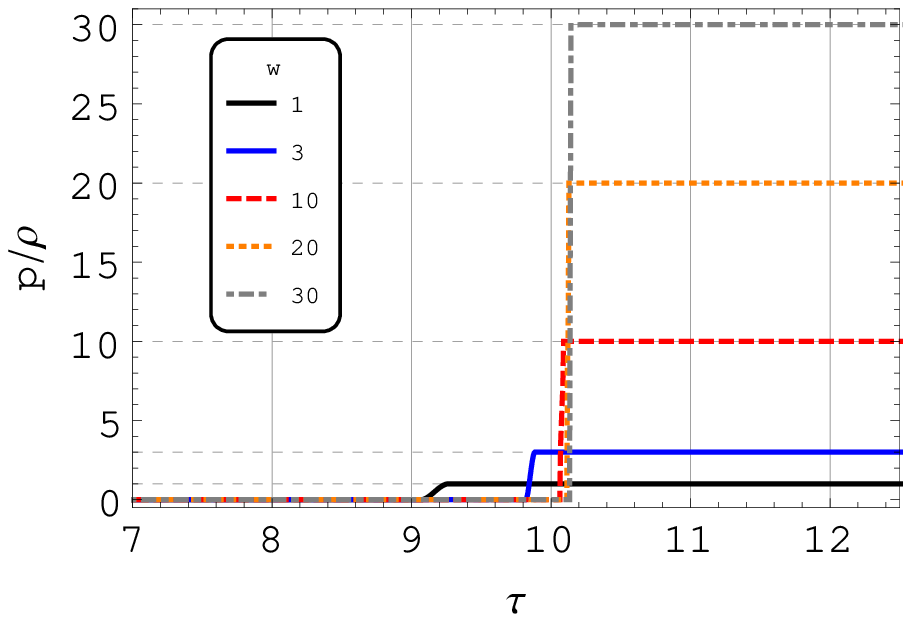}}
\end{tabular}
\caption{\label{fg:eos_pwrho_hybrid} 
The time evolution of (a) the surface density $\rho$ and (b) the ratio $p/\rho$
in the hybrid $p=w\rho$ model for $w=1,3,10,20$, and $30$. The surface
	 density $\rho$
	 monotonically increases and then
	 approaches constant as the shell slows down to the final static radius.
For each value of $w$, the ratio $p/\rho$ is kept $0$ first and then rapidly
	 shifted to $w$ at some moment.
Then, the ratio $p/\rho$ is kept $w$ after the transition.}
        \end{center}
\end{figure}

Let $V_w(R)$ be the potential of a shell with $p=w\rho$ derived in the
previous section and $V_0(R)$ be the potential of a dust shell 
falling from a static state at $R=\infty$.
For our purpose, we connect $V_w(R)$ and $V_0(R)$ with an interpolating
function $V_{{\rm smooth}}(R)$ as 
\begin{align}
 V(R) = \begin{cases}
    V_w(R):=1-M/R-m_f^2M^{2+4w}/(4R^{2+4w})-R^{4w}/(m_f^2M^{4w}) & (R\leq R_s-l) \\
    V_{{\rm smooth}}(R) & ( R_s-l \leq R \leq  R_s+l) \\
    V_{0}(R):=-M/R-M^2/(4R^2) & (R \geq R_s+l)
  \end{cases},
  \label{combined-potential}
\end{align}
where $V_{{\rm smooth}}(R)$ is a quintic function interpolating between $V_w$ and $V_0$ so that $V$, $V'$ and $V''$ are continuous.
$R_s$ is defined through $V_w(R_s)=V_0(R_s)$ so that it
corresponds to the intersection between $V_w(R)$ and $V_0(R)$.
Then, $R_{s}$ is obtained as $R_s=m_f^{1/(2w)}M$.
$l$ determines the domain of the interpolating function $V_{{\rm
smooth}}(R)$ and is chosen as $l=(R_s-R_f)/A$ with a constant $ A\geq 2$. 
\footnote{We have numerically 
calculated the time evolution of the power of
radiation for different values of $A$ in the range $2\le A\le 10$ and 
found that the difference in the obtained result is very little 
under the two hybrid potentials given by \eq{combined-potential} and
\eq{quadratic-potential}, except for that the larger value 
of $A$ caused numerical
difficulty in the latter case. So, for the numerical results presented
in this paper, 
we fix $A$ to $A=10$ and $A=2$ for the potentials
\eq{combined-potential} and
\eq{quadratic-potential}, respectively.}
We note that because the shell falls from the static state at
$R=\infty$ in this effective potential, it
emerges from the past timelike infinity as depicted in \fig{fig-null-timelike} (b).
From Eqs.~(\ref{JUNCTION1}) and (\ref{JUNCTION2}), 
the surface density $\rho$ and the ratio $p/\rho$
for different values of $w$ were numerically calculated and 
are summarized in
\fig{fg:eos_pwrho_hybrid} 
for $w=1, 3, 10, 20$ and $30$. 
One can see that in each case, as $\rho$ monotonically increases, 
$p/\rho$ is kept $0$
first, jumps from $0$ to $w$ very rapidly at some moment,
and then is kept $w$ afterwards. This is because
this model is identical to the dust shell for $R\ge R_{s}+l$, 
the pure $p=w\rho$ shell model for $R<R_{s}-l$, 
and something smoothly connecting these two regimes for $R_{s}-l\le R<R_{s}+l$
 as discussed in Sec.~\ref{section-linear}.

\subsection{Hybrid quadratic model}\label{section-quadratic}
Inspired from the hybrid $p=w\rho$ potential in the previous subsection, 
we next consider the following potential:
\begin{align}
 V(R) = \begin{cases}
    V_q(R):=-\sigma^2(R-R_f)^2/M^2 & (R\leq R_s-l) \\
    V_{{\rm smooth}}(R) & (R_s-l \leq R \leq  R_s+l) \\
    V_{0}(R)=-M/R-R^2/(4M^2) & (R \geq R_s+l)
  \end{cases},
  \label{quadratic-potential}
\end{align}
where $V_{{\rm smooth}}(R)$ is, as in the hybrid $p=w\rho$ 
potential model, a quintic function smoothly interpolating 
$V_q(R)$ and $V_0(R)$. 
In other words, the whole potential is obtained by connecting the
quadratic function and the dust potential with an interpolating
function. 
Since $R_{f}$ and $\sigma$ are related to $\epsilon$ and $\beta$ through
$R_{f}=2M(1+\epsilon^{2})$ and $\sigma^{-1}=2M\epsilon^{2\beta}$,
we can instead regard $\epsilon$ and $\beta$ as free parameters.
The larger value of $\sigma$ with fixed $R_{f}$ 
implies the larger value of $\beta$ with fixed $\epsilon$, 
corresponding to more squashed quadratic potential 
around $R=R_{f}$. 

\begin{figure}
        \begin{center}
\begin{tabular}{cc}
          \subfigure[]{\includegraphics[width=0.45\textwidth]{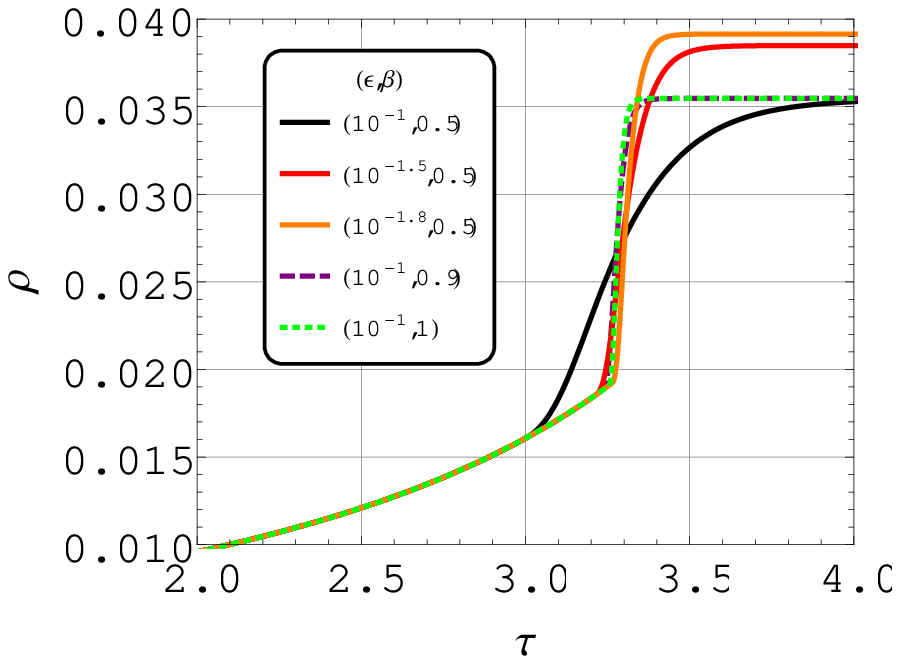}}
 & 
          \subfigure[]{\includegraphics[width=0.45\textwidth]{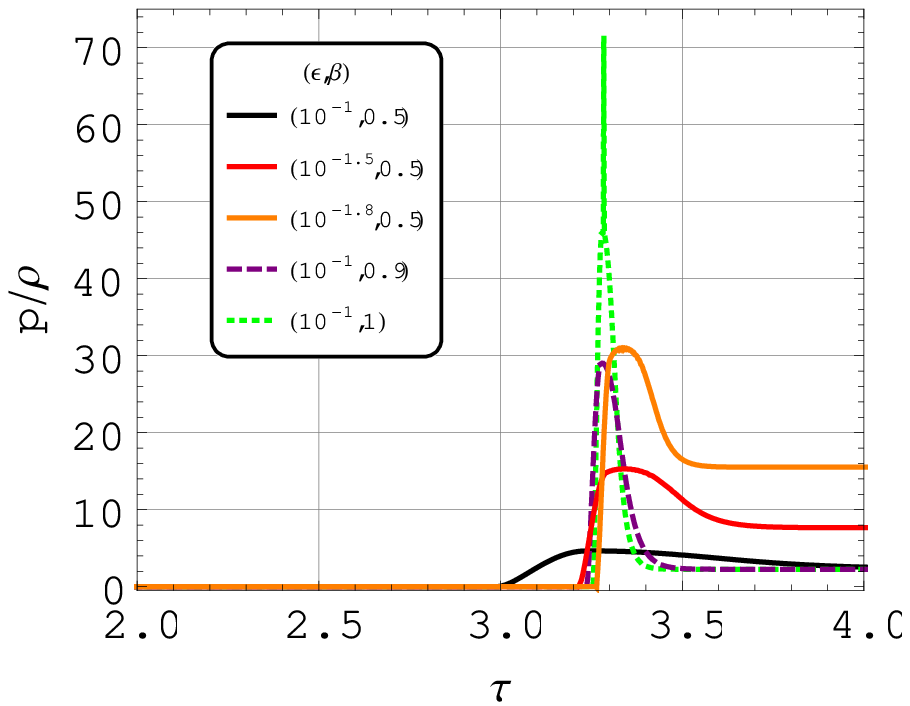}}
\end{tabular}
\caption{\label{EOS-transition-epsilon-beta} 
The time evolution of (a) the surface density $\rho$ and (b) the ratio $p/\rho$
in the hybrid quadratic model for
	 $(\epsilon,\beta)=(10^{-1},0.5)$, $(10^{-1.5},0.5)$,
	 $(10^{-1.8},0.5)$, $(10^{-1},0.9)$, and $(10^{-1},1)$. The surface
	 density $\rho$
	 monotonically increases and then
	 approaches constant as the shell slows down to the final static radius.
	 The final static radius is determined by $\epsilon$, while the
	 time scale of the transition is determined by
	 $\sigma^{-1}=2M\epsilon^{2\beta}$. For each case, 
the ratio $p/\rho$ is kept $0$ first and then rapidly
	 increased to some maximum value at some moment. Then, it 
	 turns to decrease and approaches the final value determined by
	 $\epsilon$.}
        \end{center}
\end{figure}

The time evolution of the surface density $\rho$ and the ratio $p/\rho$ 
for $(\epsilon,\beta)=(10^{-1},0.5)$, $(10^{-1.5},0.5)$,
	 $(10^{-1.8},0.5)$, $(10^{-1},0.9)$, and $(10^{-1},1)$
is shown in
\fig{EOS-transition-epsilon-beta}.
The surface density $\rho$ follows its evolution of the dust shell and then 
shifts to that of the shell with $p=w\rho$.
The final value is determined by $\epsilon$, while the time scale of the 
transition is determined by $\sigma^{-1}=2M\epsilon^{2\beta}$. 
The ratio $p/\rho$ begins with zero, increases
to a maximum value, turns to decrease, and ends with 
some positive value determined by $\epsilon$.
The maximum value of the ratio is larger for the smaller value of
$\epsilon $ and the larger value of $\beta$, 
while the final value is larger for the
smaller value of $\epsilon$ but independent from $\beta$.
All of the EOSs shown in both 
Figs.~\ref{EOS-transition-epsilon-beta} (a) and (b) are thermodynamically
unstable in the regime where $d p/d \rho <0$. 

\section{Particle creation: a numerical analysis}
\label{sec:numerical}

\subsection{Hybrid $p=w\rho$ transmissive model
\label{subsec:linearEOS_transmissive}}

Using Eqs.~(\ref{eq:P_kappa}) and (\ref{eq:kappa_transmissive}),
we numerically calculated $\kappa(u)$ and then
$P(u)$ with $\delta=0$ for each dynamical model with the transmissive
boundary condition.
We present the numerical result in the hybrid models introduced in Sec.~\ref{sec:concrete_models}.

\begin{figure}[htbp]
  \begin{center}
    \begin{tabular}{cc}
          \subfigure[]{\includegraphics[width=0.4\textwidth]{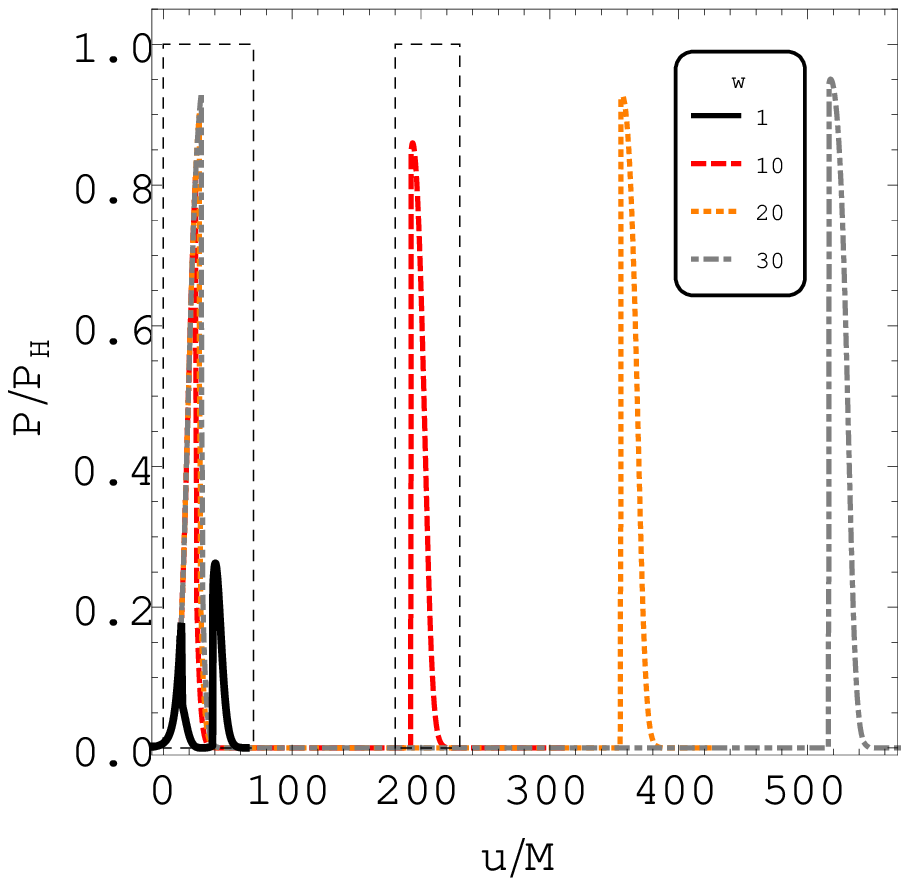}}&
          \subfigure[]{\includegraphics[width=0.4\textwidth]{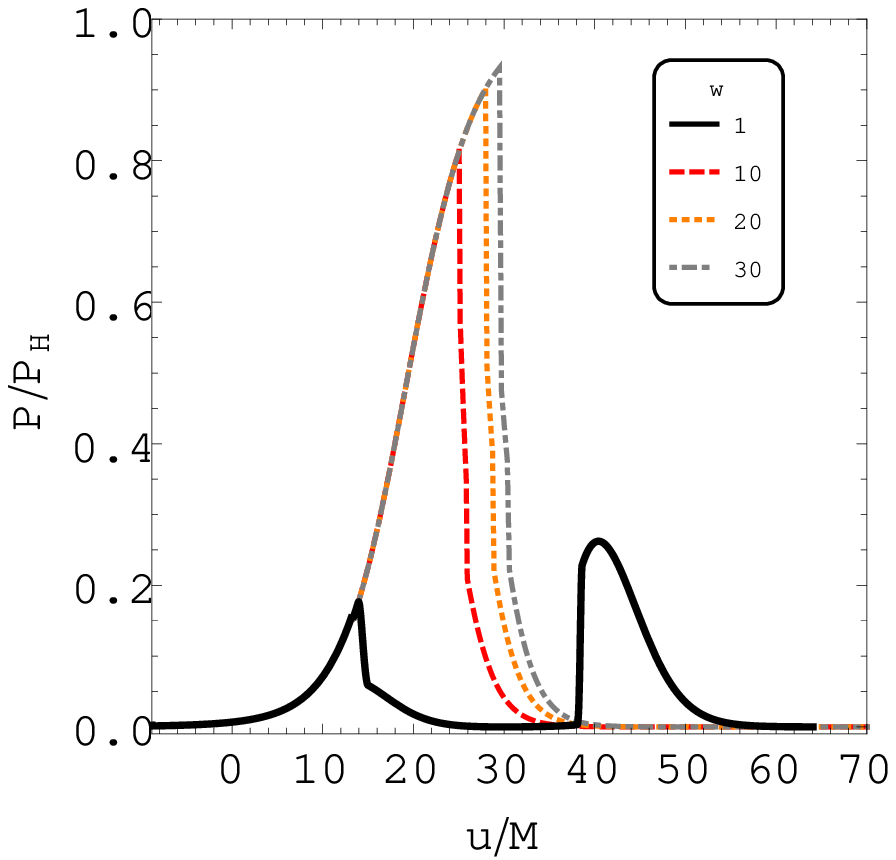}}\\
          \subfigure[]{\includegraphics[width=0.4\textwidth]{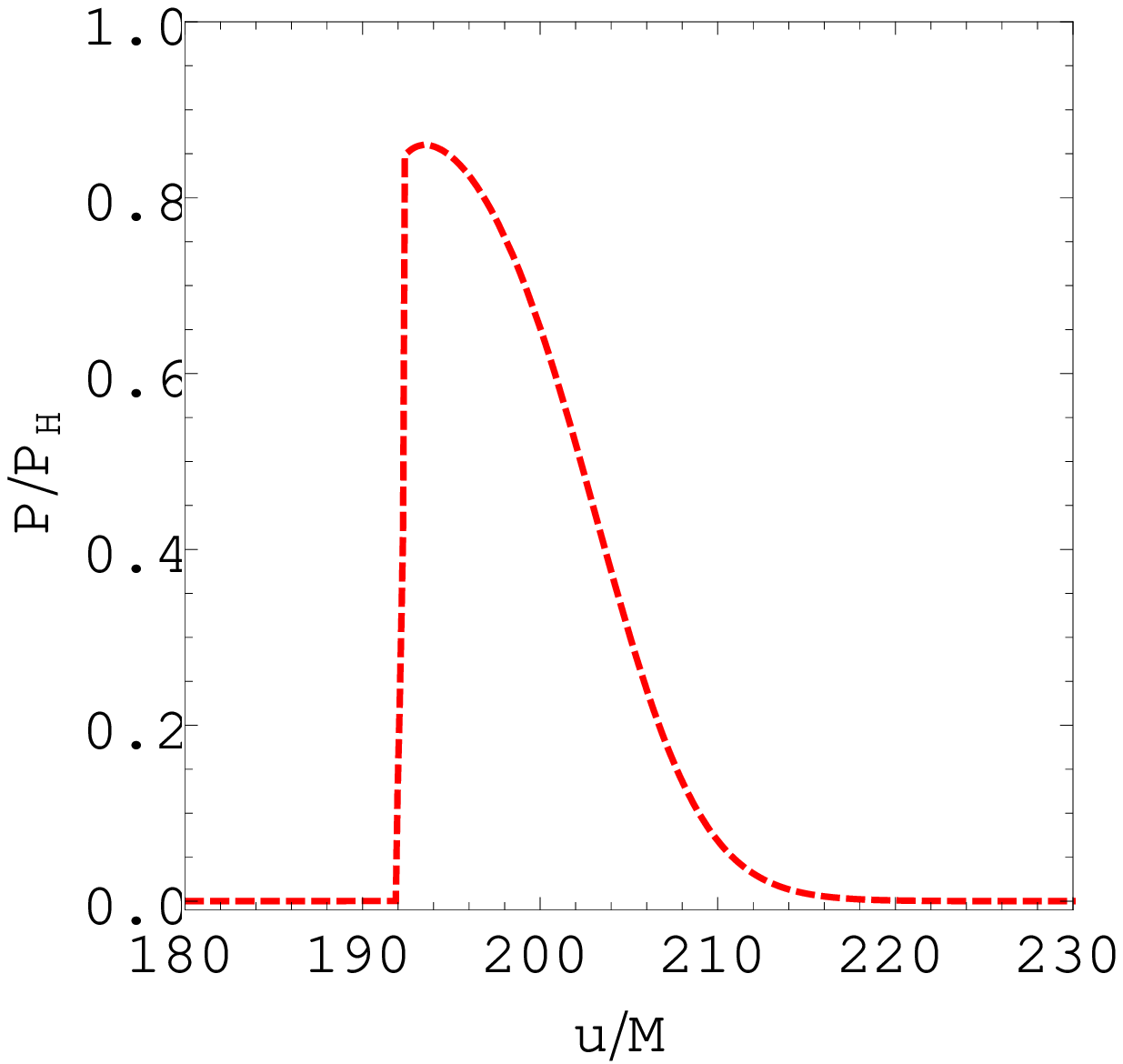}}&
          \subfigure[]{\includegraphics[width=0.4\textwidth]{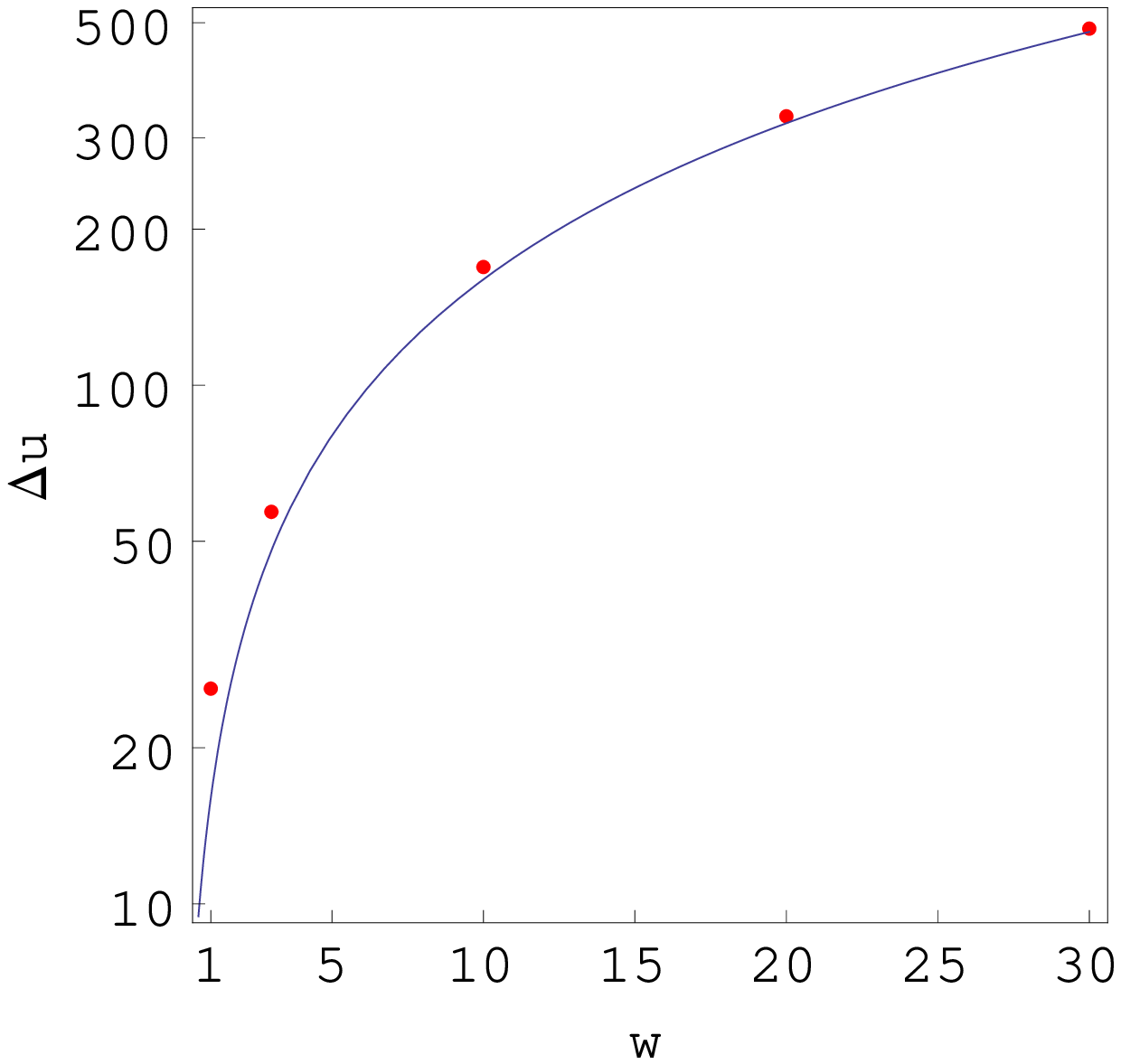}}
    \end{tabular}
\caption{The power of radiation in the hybrid
   $p=w\rho$ model with the transmissive boundary condition. (a) shows the time evolution of the power for $w=1,10,20$, and $30$.
   (b) is the enlargement of the first-peak region indicated by the left
   rectangle of dashed lines in (a). (c) is the enlargement of
   the second-peak region for $w=10$ indicated by the right rectangle 
   of dashed lines in (a).
   (d) shows the relation between $\Delta u$ and $w$ 
   for $w=1,3,10,20$, and $30$, where
   $\Delta u$ is the time interval of the two peaks. The analytic
   formula $\Delta u= 16wM$ is also plotted with a solid line in (d).
 \label{fig-combined-potential-radiation}}
  \end{center}
\end{figure}

Figure~\ref{fig-combined-potential-radiation} shows the numerical result
for the power of radiation in the hybrid $p=w\rho$ model with the
transmissive boundary condition.
Figure~\ref{fig-combined-potential-radiation} (a) shows the time
evolution of the power for $w=1, 10, 20$, and $30$. 
The horizontal axis denotes $u$ normalized by $M$.
The power plotted are normalized by the power of the standard Hawking
radiation $P_H:=(48\pi)^{-1}(16M^2)^{-1}$.
From this figure, we find that there are two major peaks for each value of
$w$.
The first peak is emitted immediately after the shell gets very close to the
final radius, while the second peak is much later.
Figures~\ref{fig-combined-potential-radiation} (b) and (c) are the 
enlarged figures of the first peaks for the different values of $w$
and of the second peak for $w=10$, respectively, where the enlarged
areas are indicated
with the rectangles of dashed lines in
Fig.~\ref{fig-combined-potential-radiation} (a).
In Fig.~\ref{fig-combined-potential-radiation} (b), we can see that the power first gradually increases in time,
which corresponds to the rise of THR. 
Then, this rise of THR is suddenly interrupted before it 
reaches $P_{H}$. 
The larger the value of $w$, the later the time of the
interruption. 
No significant bump is seen after the interruption of THR in
Fig.~\ref{fig-combined-potential-radiation} (b)
except for $w=1$, for which the second peak already appears in
Fig.~\ref{fig-combined-potential-radiation} (b).
In Fig.~\ref{fig-combined-potential-radiation} (c), we can see the time
evolution of the power at the second peak for $w=10$. 
The power is first negligibly small after the first peak. Then, it
suddenly rises nearly to $P_{H}$ very quickly and then decays in the
time scale of a few
tens of $M$. This time profile of the second peak 
is common to the different values of $w$.
The second peak is associated with negative $\kappa$ according to the
qualitative analysis in Paper I.
We observe that the peak values increase as $w$ is
increased, although $P$ does not exceed $P_{H}$ and remains 
as strong as $P_{H}$ at most even for the larger values of $w$.
The time interval between the two radiation peaks, $\Delta u$, increases as
$\epsilon$ is decreased. 
In Fig.~\ref{fig-combined-potential-radiation} (d), 
we plot the relation between the time
interval of the two peaks, $\Delta u$,
and $w$. The time interval $\Delta u$ increases 
as $w$ is increased.
We also plot the relation $\Delta u\simeq 16wM$ with a
solid line in the same figure.
This relation can be derived from 
$\Delta u\simeq 4M/\epsilon$ obtained in Paper I~\cite{Harada:2018zfg}
and $\epsilon = 1/(4w)$.
We can see that the numerical result is in good agreement with the
relation $\Delta u\simeq 16 w M$.

\subsection{Hybrid quadratic transmissive model
\label{subsec:quadratic_transmissive}}
\begin{figure}[htbp]
  \begin{center}
    \begin{tabular}{cc}
          \subfigure[]{\includegraphics[width=0.4\textwidth]{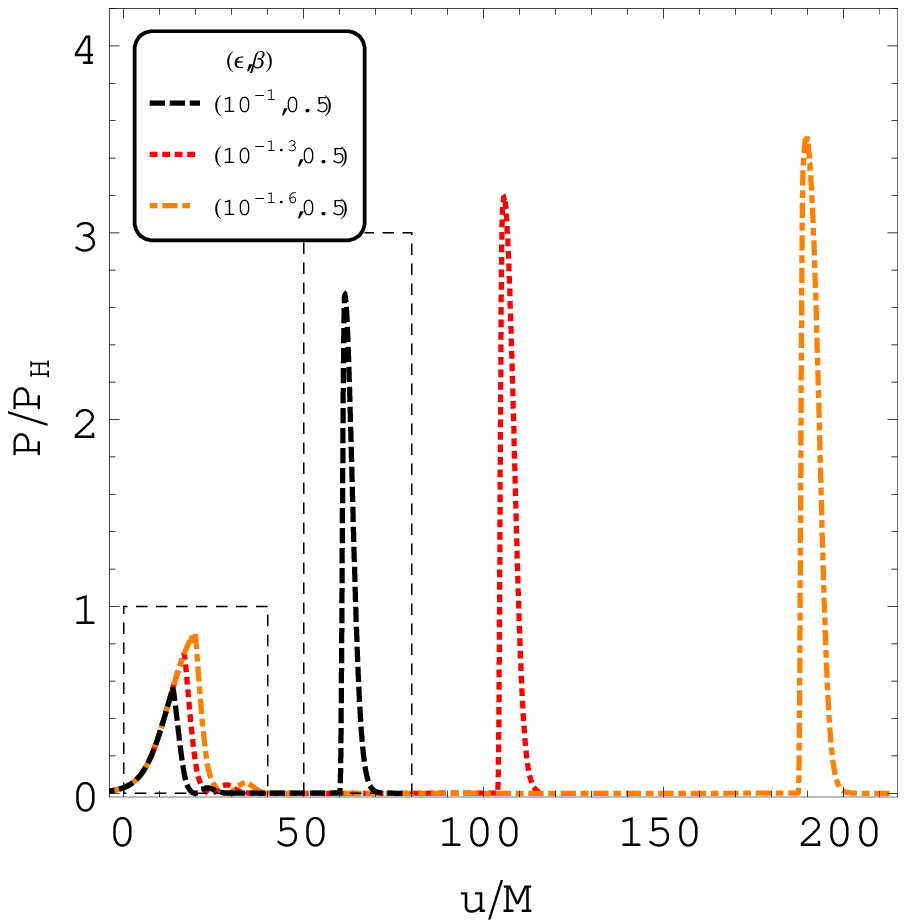}}&
          \subfigure[]{\includegraphics[width=0.4\textwidth]{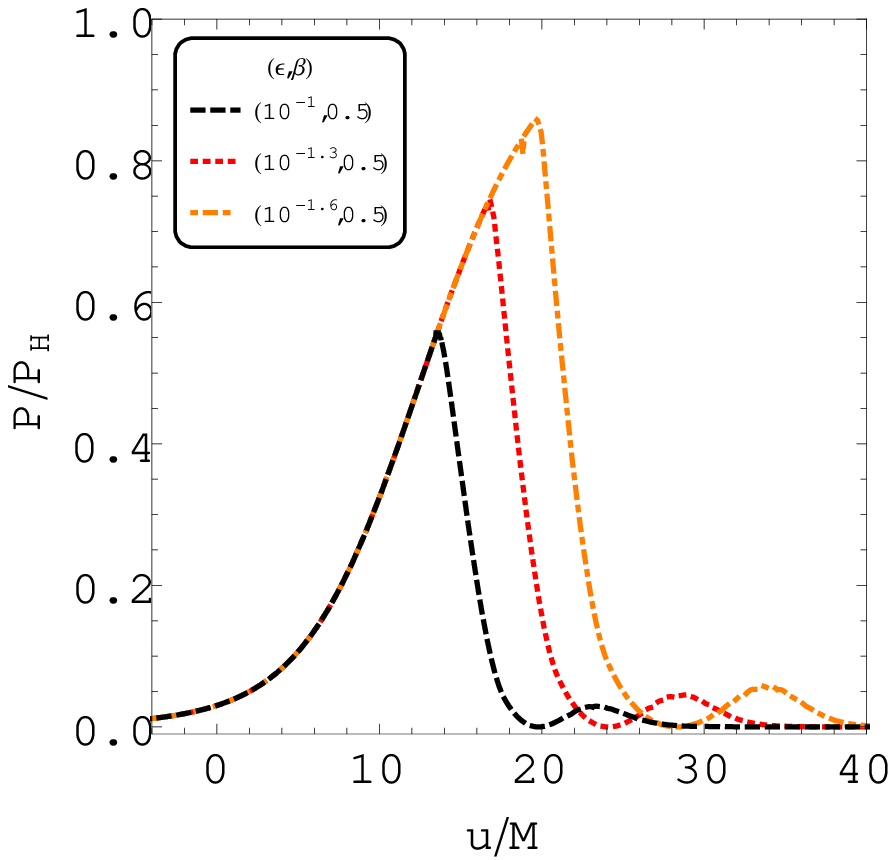}}\\
          \subfigure[]{\includegraphics[width=0.4\textwidth]{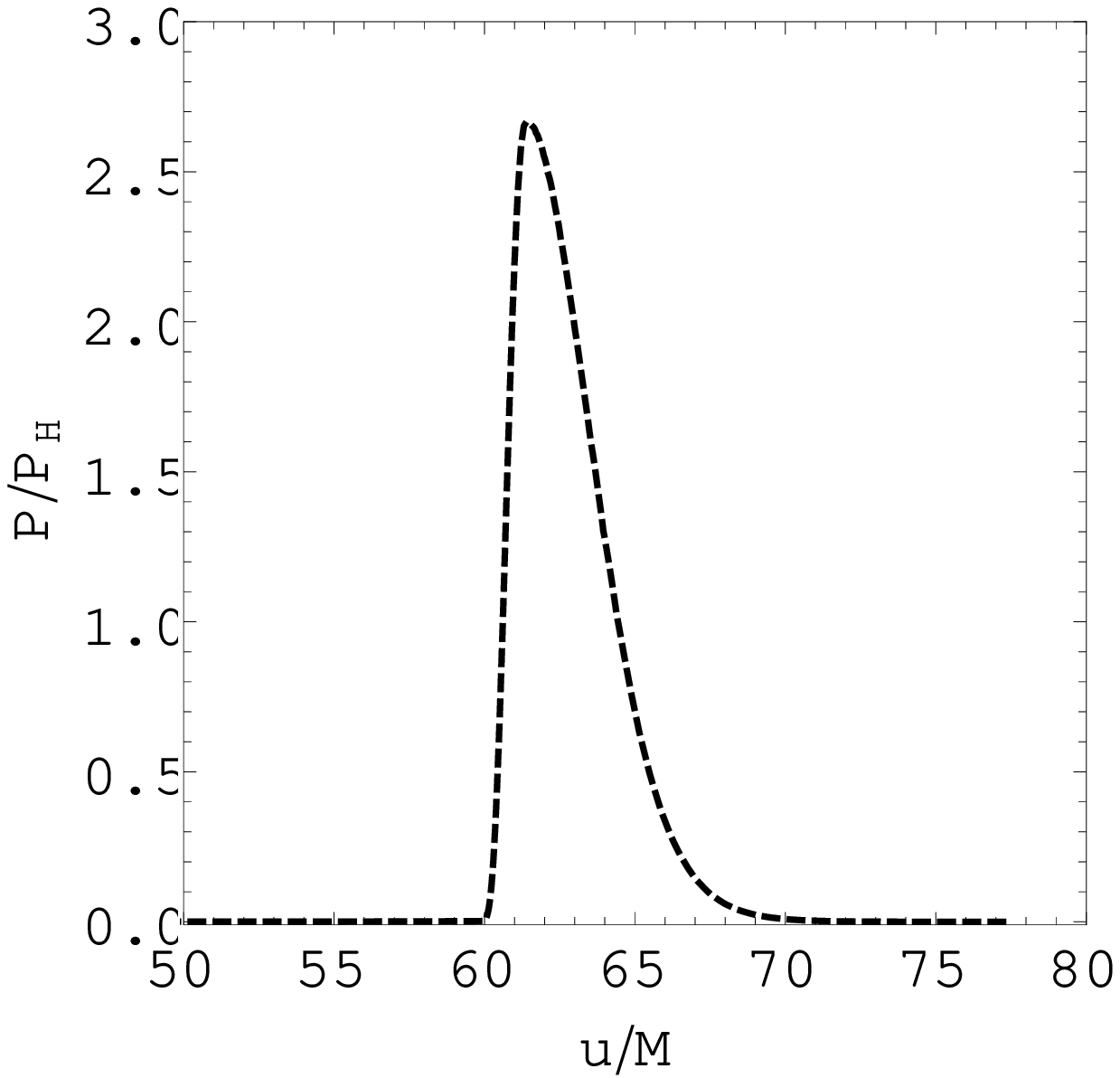}}&
          \subfigure[]{\includegraphics[width=0.4\textwidth]{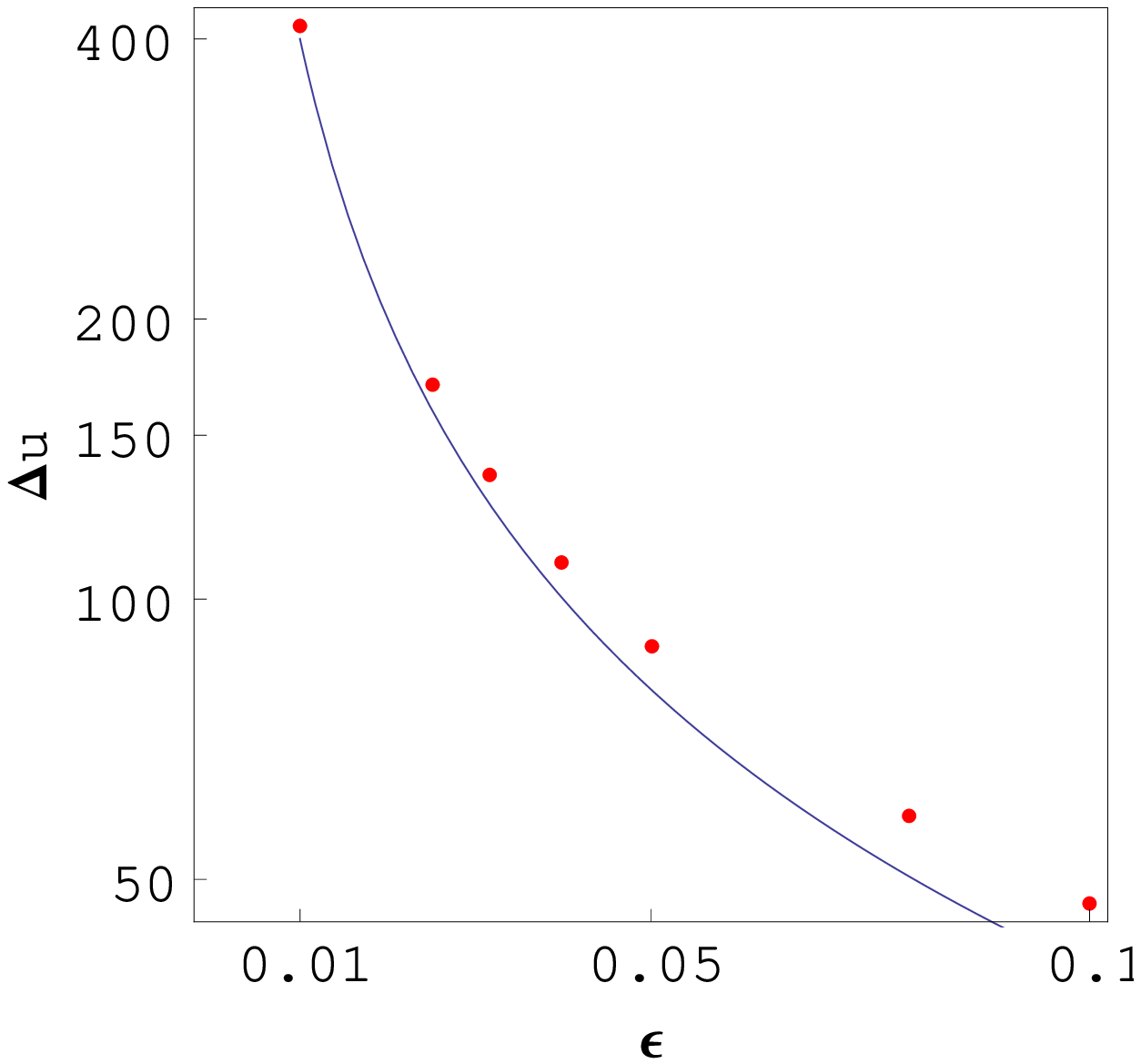}}
    \end{tabular}
\caption{
   The power of radiation in the hybrid
   quadratic model with the transmissive boundary condition:
   $\epsilon$-dependence. (a) shows the time evolution of the power for
   $\epsilon=10^{-1}, 10^{-1.3}$, and $10^{-1.6}$, where $\beta=0.5$ is fixed. 
   (b) is the enlargement of the first-peak region indicated by the left
   rectangle of dashed lines in (a). (c) is the enlargement of
   the second-peak region indicated by the right rectangle 
   of dashed lines in (a) for $(\epsilon,\beta)=(10^{-1},0.5)$ .
   (d) shows the relation between $\Delta u$ and $\epsilon$ 
   for different values of $\epsilon$, where $\beta=0.5$ is fixed. 
   The analytic formula $\Delta u= 4M/\epsilon$ is plotted with 
   a solid line in (d).
 \label{fig-quadratic-transmissive-epsilon}}
  \end{center}
\end{figure}

\begin{figure}[htbp]
  \begin{center}
    \begin{tabular}{cc}
          \subfigure[]{\includegraphics[width=0.4\textwidth]{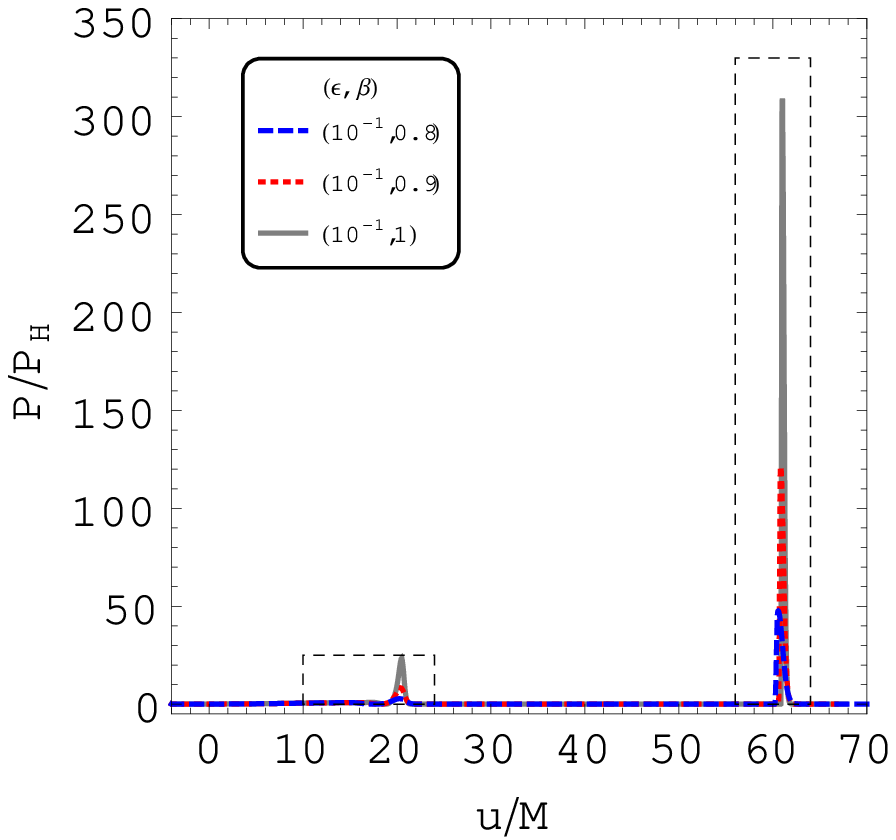}}&
          \subfigure[]{\includegraphics[width=0.4\textwidth]{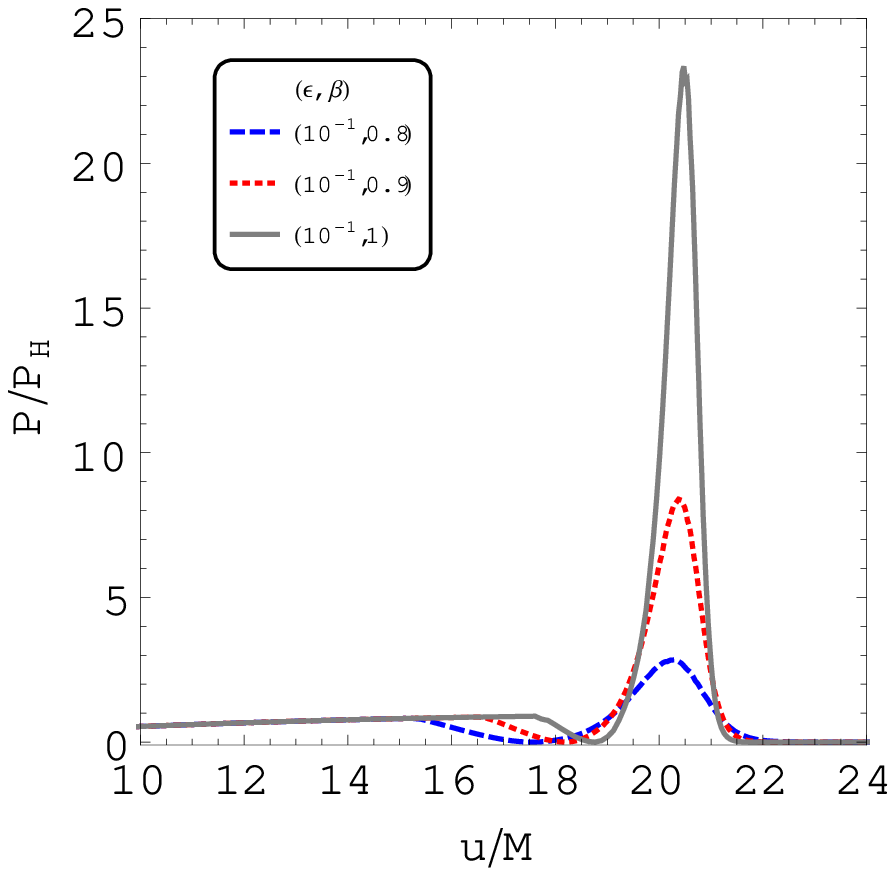}} \\
          \subfigure[]{\includegraphics[width=0.4\textwidth]{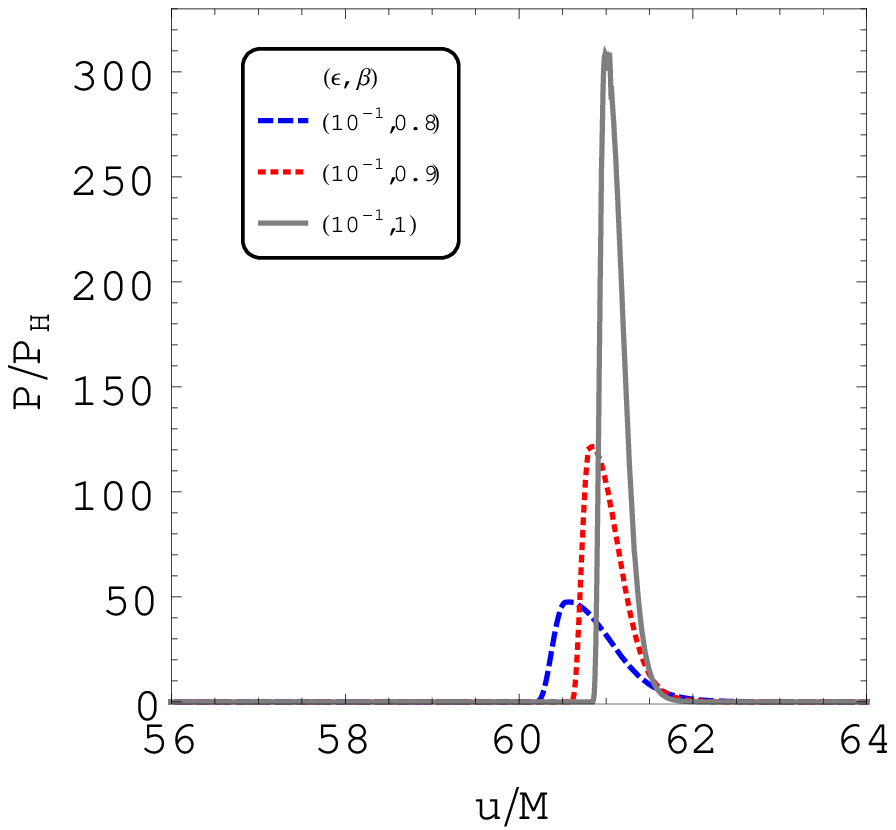}}&
          \subfigure[]{\includegraphics[width=0.4\textwidth]{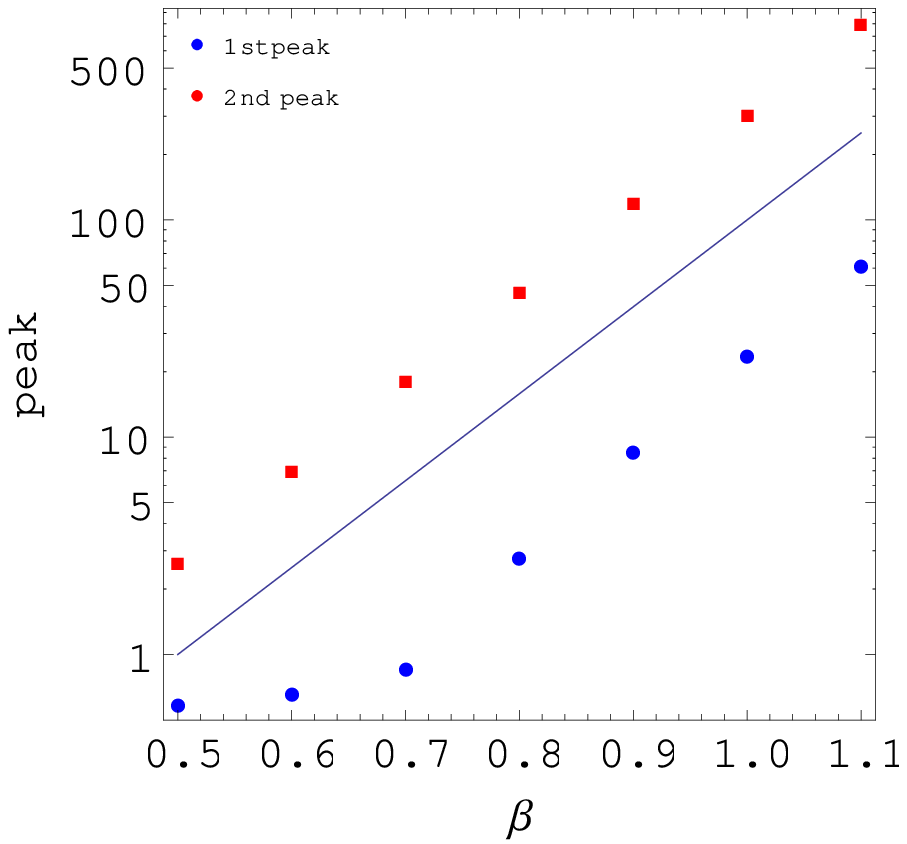}}
    \end{tabular}
\caption{ 
   The power of radiation in the hybrid
   quadratic model with the transmissive boundary condition:
   $\beta$-dependence. (a) shows the time evolution of the power for
   $\beta=0.8, 0.9$, and $1.0$, where $\epsilon=10^{-1}$ is fixed. 
   (b) is the enlargement of the first-peak region indicated by the left
   rectangle of dashed lines in (a). (c) is the enlargement of
   the second-peak region indicated by the right rectangle 
   of dashed lines in (a).
   (d) shows the relation between the values of the two peaks normalized
   by $P_{H}$ and $\beta$ with $\epsilon=0.1$. 
   The analytic formula $P/P_{H}=\epsilon^{2(1-2\beta)}$ is plotted with 
   a solid line in (d).
 \label{fig-quadratic-transmissive-beta}}
  \end{center}
\end{figure}

Figures \ref{fig-quadratic-transmissive-epsilon} and 
\ref{fig-quadratic-transmissive-beta} show the numerical result in the hybrid quadratic potential
given by \eq{quadratic-potential}. Recall that the 
two parameters of this potential
$R_{f}$ and $\sigma$ can also be parameterize by two nondimensional
parameters $\epsilon$ and $\beta$ 
with $R_{f}=2M(1+\epsilon^{2})$ and $\sigma^{-1}=2M\epsilon^{2\beta}$.

First, we see the $\epsilon$-dependence of the power of radiation in Fig.~\ref{fig-quadratic-transmissive-epsilon}.
Figure \ref{fig-quadratic-transmissive-epsilon} (a) shows the time
evolution of the power of radiation for 
$\epsilon=10^{-1}, 10^{-1.3}$, and $10^{-1.6}$, where $\beta=0.5$ is fixed.
We can see that there are two bursts, the first one exceeded by the
second one. The peak values of the both increase as $\epsilon$ is
decreased, although they still remain of the order of $P_{H}$.
Figures \ref{fig-quadratic-transmissive-epsilon} (b) and (c) are the 
enlarged figures of the first peaks for the different values of $\epsilon$
and of the second peak for $\epsilon=10^{-1}$, respectively, where the enlarged
areas in Figure~\ref{fig-quadratic-transmissive-epsilon} (a) 
are indicated with the rectangles of dashed lines.
In Fig.~\ref{fig-quadratic-transmissive-epsilon} (b), we can see that 
the rise of THR is suddenly interrupted before it reaches $P_{H}$. 
The smaller the value of $\epsilon$, the later the time of the interruption.
Unlike in the hybrid $p=w\rho$ model, we can see a small bump after the
THR is interrupted. In fact, $\kappa$ first increases nearly
to the Hawking value $\kappa_{H}$, then decreases, crosses zero, and
becomes negative. The negative value of $\kappa$ is due to the braking
of the collapse. This is discussed in Paper I, although this fine
structure is omitted in the schematic figures, Figs. 5 and 6 in 
Paper I~\cite{Harada:2018zfg}. 
As the brake ceases, the value of $\kappa$ approaches zero.  
In Fig.~\ref{fig-quadratic-transmissive-epsilon} (c), we can see the time
evolution of the power at the second peak. Its feature is similar to
that in the hybrid $p=w\rho$ model except for that the value of the
second peak is significantly larger than $P_{H}$. 
The peak value is larger and the time scale of the decay is longer for
the smaller value of $\epsilon$.
The second peak is associated with negative $\kappa$ according to the
qualitative analysis in Paper I~\cite{Harada:2018zfg}.
The interval between the two
bursts is plotted as a function of $\epsilon$ for fixed $\beta=0.5$ in
Fig.~\ref{fig-quadratic-transmissive-epsilon} (d), from which we 
can see that the numerical
result agrees well with the relation $\Delta u=4M/\epsilon$ derived by
the qualitative analysis in Paper I~\cite{Harada:2018zfg}.  

Next, we see the $\beta$-dependence shown in 
\fig{fig-quadratic-transmissive-beta}.
In \fig{fig-quadratic-transmissive-beta} (a), we plot the numerical
result for the time evolution of the power for 
different values of $\beta$, while $\epsilon=0.1$ is fixed.
This corresponds to changing $\sigma$ while fixing $R_{f}$.
In this case, the situation drastically changes from the case of
$\beta=0.5$. We can see that the values of the both peaks can be 
much greater than $P_{H}$ for $\beta>0.5$.
The larger the value of $\beta$, the larger both the first peak and second
peak.
Figures \ref{fig-quadratic-transmissive-beta} (b) and (c) 
are the enlarged figures of the first peaks and second peaks,
respectively, where the enlarged
areas in Fig.~\ref{fig-quadratic-transmissive-beta} (a) 
are indicated with the rectangles of dashed lines.
In Fig.~\ref{fig-quadratic-transmissive-beta} (b), we can see that 
the THR is suddenly interrupted and the power decreases to zero. 
Then, the power turns to increase
and reaches the first peak, which is much larger than $P_{H}$.
This first peak is associated with negative $\kappa$ 
due to the braking of the collapse. After the first peak, the power
soon decays to a negligibly small value. 
In Fig.~\ref{fig-quadratic-transmissive-beta} (c), we can see the 
detailed time evolution of the second peak, which is associated with
negative $\kappa$. 
Figure~\ref{fig-quadratic-transmissive-beta} (d) shows
the relation between the values of the two peaks
and $\beta$ for fixed $\epsilon=0.1$. From this figure,  
we can see that the first and the second peaks are proportional to
$P/P_H=\epsilon^{2(1-2\beta)}= (\epsilon\cdot 2M\sigma)^{2}$ 
for $\beta\ge 0.5$, 
which is derived as Eq. (6.12) in Paper I~\cite{Harada:2018zfg}
and plotted by a solid line in \fig{fig-quadratic-transmissive-beta} (d). 
We can see that $\beta$ together with $\epsilon$ 
controls the strength of the peak power.
The peak value becomes larger for the smaller value of $\epsilon$
and the larger value of $\beta$.
Within our numerical result, the strongest burst was $P/P_H\simeq 800$ 
as the second peak for $(\epsilon,\beta)=(0.1,1.1)$.
We can see from \fig{fig-quadratic-transmissive-beta} (a) 
that the time interval between the two peaks for the different values of
$\beta$ with fixed $\epsilon$ is almost constant, which is consistent 
with the general relation $\Delta u=4M/\epsilon$.

To summarize, the closer the final static radius to the horizon
radius, the longer the interval between the first and second bursts, 
while the more squashed the parabolic potential or the stronger the
brake, the stronger both the first and second bursts.

\subsection{Hybrid $p=w\rho$ reflective model}
\label{section-reflective}
Using Eqs.~(\ref{eq:P_kappa}) and
(\ref{eq:kappa_reflective}), we numerically calculated $\kappa(u)$ and then
$P(u)$ with $\delta=0$ for each dynamical model with the reflective
boundary condition.
Here, we present the numerical result for the power 
evolution emitted from a collapsing reflective surface.
Note that to calculate the power of radiation, we do not have to 
specify the internal structure of the collapsing body but only the
dynamics of the reflective surface. 
Here, however, just for convenience,
we continue to use the same terminology to refer to the 
dynamics of the reflective surface as if it is a hollow spherical 
shell.

\begin{figure}
        \begin{center}
          \includegraphics[width=0.4\textwidth]{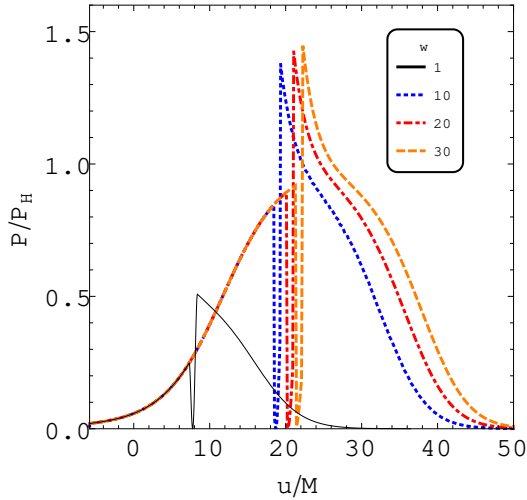} 
           \caption{The time evolution of the power of radiation in the 
	 hybrid $p=w\rho$ potential with the reflective boundary
	 condition for $w=1$, $10$, $20$, and $30$. \label{fig-reflective-linearEOS}}
        \end{center} 
\end{figure}

Figure \ref{fig-reflective-linearEOS} shows the numerical result for 
the power evolution in the hybrid $p=w\rho$ model
with $w=1, 10, 20$, and $30$ for the reflective boundary condition.
We can clearly see that the THR first rises and is suddenly interrupted
to zero in a time scale much shorter than $M$.  
The THR is nearly complete for $w\gtrsim 10$, while it is far from complete
for $w=1$. After the THR is interrupted in the very short time scale,
we can see a burst arises in the same short time scale to 
the value as strong as $P_{H}$.  
This post-Hawking burst is associated with negative $\kappa$ and 
decays in the time scale of a few tens of $M$. 
The time evolution is delayed for larger values of $w$ but 
it is not very sensitive for $w\gtrsim 10$. This is consistent
with the qualitative analysis showing the dependence 
$\sim \log \epsilon^{-1}\sim \log w$ of the duration of the THR 
in terms of $u$ as is 
discussed in Sec.~\ref{sec:standard_collapse}.  
The burst is single and there is no further burst.
This is in contrast to the transmissive case, where such a burst does not appear immediately after the THR
as discussed in Sec.~\ref{subsec:linearEOS_transmissive}.

\subsection{Hybrid quadratic reflective model}

\begin{figure}[htbp]
  \begin{center}
    \begin{tabular}{c}

      \begin{minipage}{0.5\hsize}
        \begin{center}
          \subfigure[]{\includegraphics[width=0.7\textwidth]{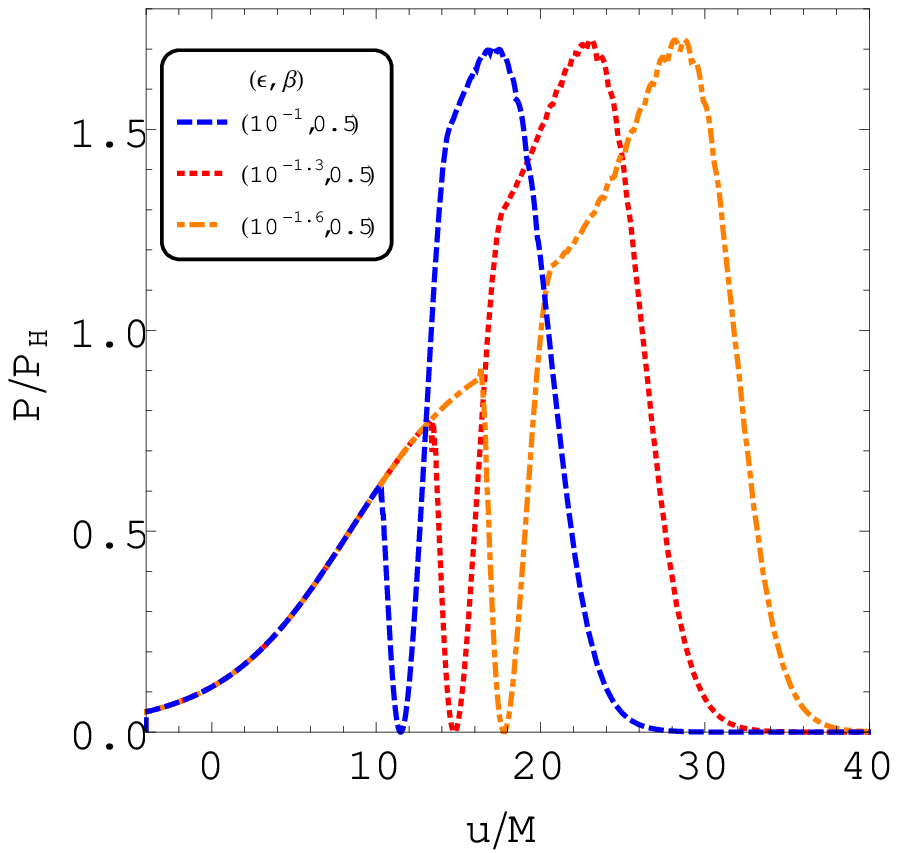}}
        \end{center}
      \end{minipage}
      
      \begin{minipage}{0.5\hsize}
        \begin{center}
          \subfigure[]{\includegraphics[width=0.7\textwidth]{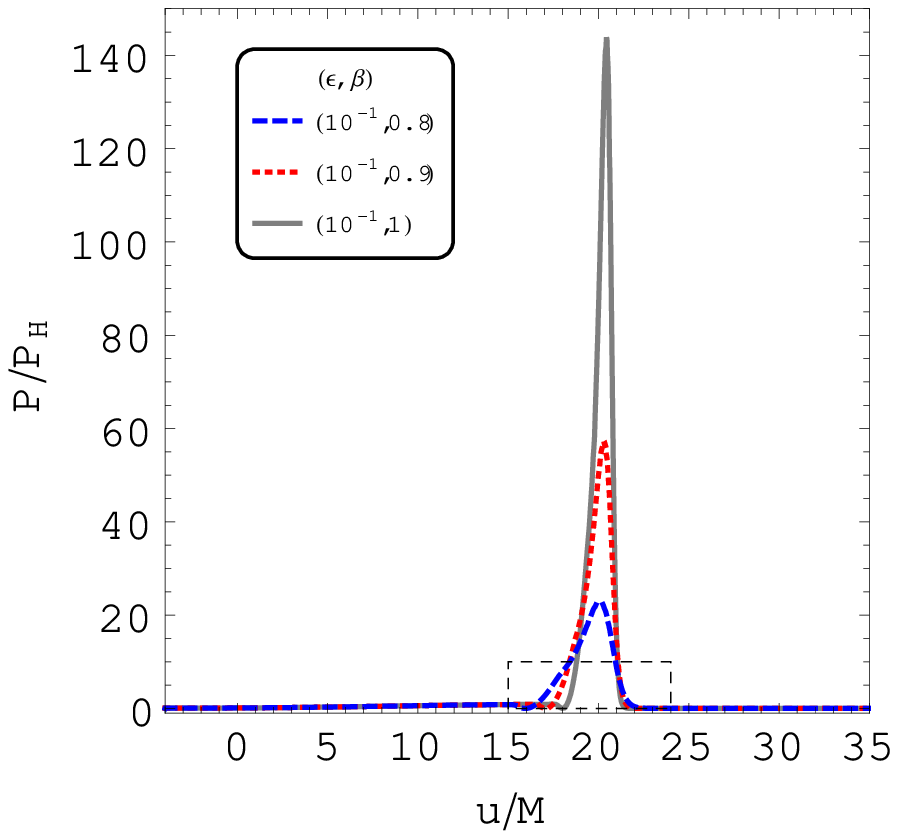}}
        \end{center}
      \end{minipage}

\\
      \begin{minipage}{0.5\hsize}
        \begin{center}
          \subfigure[]{\includegraphics[width=0.7\textwidth]{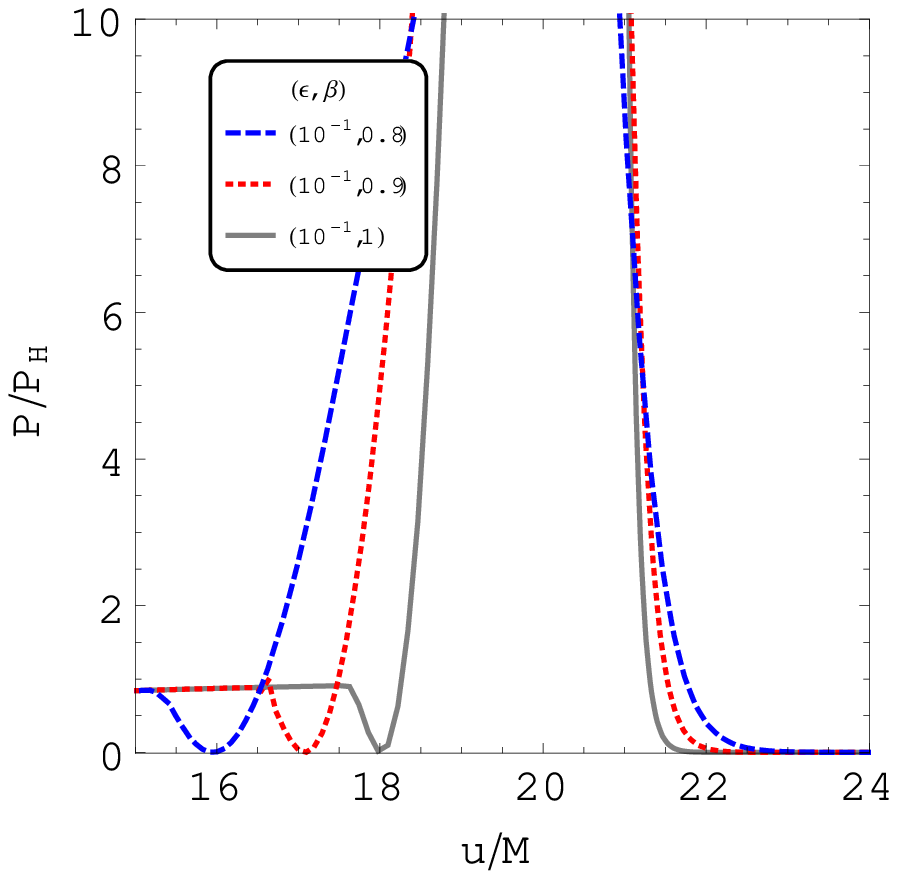}}
        \end{center}
      \end{minipage}

      \begin{minipage}{0.5\hsize}
        \begin{center}
          \subfigure[]{\includegraphics[width=0.7\textwidth]{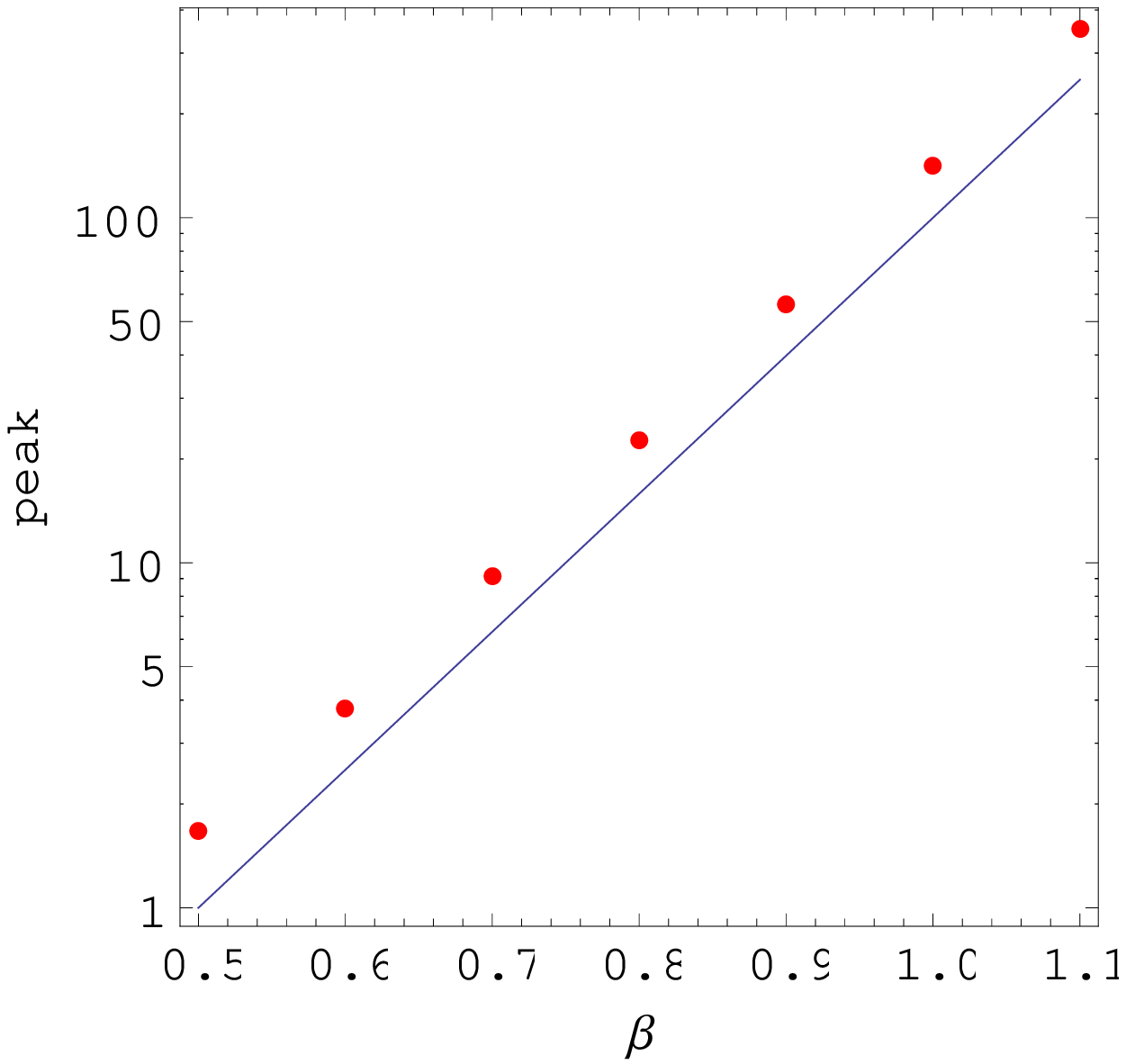}}
        \end{center}
      \end{minipage}

    \end{tabular}
\caption{The power of radiation in
the hybrid quadratic model 
with the reflective boundary condition. (a) shows 
$P(u)$ for $\epsilon=10^{-1}$, $10^{-1.3}$, and $10^{-1.6}$ with $\beta=0.5$. 
(b) shows $P(u)$ for $\beta=0.8$, $0.9$, and $1.0$ with $\epsilon=0.1$.
(c) is the enlargement of the region indicated by the rectangle of dashed
   lines in (b).
(d) shows the relation between the peak value and $\beta$ with $\epsilon=0.1$. 
The analytic formula $P/P_{H}=\epsilon^{2(1-2\beta)}$ is also plotted 
with a solid line in (d).
\label{fig-burst-reflective}}
  \end{center}
\end{figure}

Figure~\ref{fig-burst-reflective} shows the numerical result for the
power of radiation from a reflective
surface in the hybrid quadratic model.
Figure \ref{fig-burst-reflective} (a) shows the time evolution of
the power for $\epsilon=10^{-1}$, $10^{-1.3}$, and
$10^{-1.6}$ with $\beta=0.5$.
We can see that the rise of THR is suddenly interrupted in a short
time scale. A peak appears immediately after the interruption of 
THR. This post-Hawking burst is associated with negative $\kappa$ and
decays in the time scale of a few tens of $M$. The smaller the value of
$\epsilon$, the later the interruption of THR and the appearance of the post-Hawking burst.
This is in contrast to the transmissive case, where there is only a small bump immediately after the
interruption of THR as discussed in 
Sec.~\ref{subsec:quadratic_transmissive}.
Figure \ref{fig-burst-reflective} (b) shows the power evolution 
for $\beta=0.8$, $0.9$, and $1.0$ with $\epsilon=0.1$. 
We can see that the peak value is much larger than $P_{H}$ for these
values of $\beta$. 
Figure \ref{fig-burst-reflective} (c) shows the enlarged figure of the 
peaks, where the enlarged region is indicated by the rectangle of dashed
lines in Fig.~\ref{fig-burst-reflective} (b).
We can see that the THR first rises but is suddenly interrupted. 
The power drops to zero, turns to increase, rises to a peak, and then decays to zero.
The $\beta$-dependence of
the peak value for $\epsilon=0.1$ is plotted in 
Fig.~\ref{fig-burst-reflective} (d), in which the formula
$P/P_H=\epsilon^{2(1-2\beta)}=(\epsilon\cdot 2M\sigma)^{2}$, 
which is derived in 
Sec.~\ref{subsec:specific_models}, 
is also plotted by a solid line.
We can see that the numerical result 
is in good agreement with the formula. 
As in the transmissive surface, 
both $\epsilon$ and $\beta$ control the peak values of the power, 
while $\epsilon$ delays the single peak although it is not very
sensitive for $\epsilon\lesssim 10^{-1}$.

To summarize the numerical result for 
the reflective surface, we find that after the
power initially rises gradually as the THR develops, 
the power suddenly vanishes in a very short time scale.
This moment of the vanishing power corresponds to the change of 
the sign of $\kappa$ from positive 
to negative. After this interruption of THR, a 
burst arises associated with negative $\kappa$, 
exceeds $P_{H}$ and decays in the time scale of a few tens of $M$. 
This post-Hawking burst is stronger than that in the transmissive case 
for the same collapse dynamics.
On the other hand, it is interesting that the strength and the duration
of each burst in the reflective case is not so different from the first
burst in the transmissive case for the same dynamics of the shell. 
The crucial difference is that the reflective surface emits a single
burst, while the transmissive emits two.

\section{Discussion and conclusion}
\label{sec:conclusion}

We have studied particle creation in the collapse to an ultracompact
object with two different boundary conditions at the surface of the
object and with two different matter or potential models for the 
dynamics of the surface. In Paper I, the authors studied a
collapsing transmissive hollow shell.
They analytically showed the existence of 
the THR followed by a couple of bursts separated each other by the long time
interval and how the strength and duration of the bursts 
depend on the dynamics of the shell.
In the current paper, we
have also studied a collapsing reflective surface and 
analytically showed the existence of the 
THR followed by a single burst 
and how the strength and duration of the burst depend on the
dynamics of the surface.
We have found that the 
numerical result confirms the analytical result. 
Within our numerical result, the closer the final static radius to the
horizon radius, the larger the peak values of the power.  
If the EOS of the matter is described by $p=w\rho$ in its final stage of 
evolution, the power of the two bursts remain of the order of that of the
Hawking radiation at most. On the other hand, we have successfully 
constructed a concrete model described by a quadratic potential in its final
stage of evolution that emits a burst of radiation which is much stronger
than the Hawking radiation. It depends on the boundary condition on the 
surface whether the burst is single or double.
The present numerical result suggests that the EOS for the matter on 
the shell that emits bursts much stronger than the Hawking radiation
has a large maximum of the ratio $p/\rho$ implying a thermodynamically unstable 
regime in the late-time phase of evolution, although further
investigation is yet necessary to draw a general statement.

It should be noted that in the current model
the formed ultracompact object is necessarily dynamically unstable. 
This is partly because the surface cannot stay at 
the local minimum of the effective potential 
but the local maximum if the energy of the shell is conserved
during the collapse and partly because the effective potential 
of the shell connecting the Minkowski domain and the 
Schwarzschild domain cannot have a local minimum.
On the other hand, the formation of a stable 
object as a result of gravitational collapse is possible 
if the system loses energy and if the interior is the de-Sitter
spacetime. 
For example, Visser and
Wiltshire~\cite{Visser:2003ge} found that there are 
some physically reasonable EOSs for 
the transition layer that lead to stable gravastar configurations.
Nakao, Yoo and Harada~\cite{Nakao:2018knn} presented the formation scenario of a stable 
gravastar after the emission of null shells from a timelike shell. 
In this context, the current analysis of toy models 
suggests a possible mechanism for a collapsing object to lose its energy by quantum particle creation 
and thus that
a stable ultracompact object is possible as a result of gravitational collapse 
if the backreation of particle emission is taken into account and 
if a non-empty interior gives the effective potential a local minimum. 

We also discuss that the possibility of echoes in particle creation.
Since we have adopted geometric optics approximation, the particle 
emission does not show any echoes. However, if the backscattering due to 
the effective potential is taken into account, scalar waves will be 
reflected due to the potential barrier around at $r=3M$. This may
induce echoes in particle creation both in the transmissive and
reflective cases.

\vspace{1cm}

While the current work was being finalized, a preprint~\cite{Barcelo:2019eba}
appeared on the arXiv. Although the result of the current paper
partially has some overlap with Ref~\cite{Barcelo:2019eba}, 
both papers are complementary to each other in other part.

\acknowledgments

The authors are grateful to D.~Miyata, V.~Cardoso, K.~Nakashi and 
T.~Tanaka for their helpful comments. This work
was partially supported by JSPS KAKENHI Grant Number
JP19K03876 (T.H.).

\appendix

\section{Junction conditions}\label{setup-section}
A singular, spherical and timelike hypersurface $\Sigma$ (thin-shell) partitions the spacetime into the inner ($-$) and the outer ($+$) region.
Each of the metric of the solutions can be, with the coordinates $(t, r, \theta, \phi)$, written by
\begin{align}
\D s_\pm^2=-f_\pm(r_\pm)\D t^2_\pm+f_\pm(r_\pm)^{-1}\D r^2_\pm+r^2_\pm(\sin^2\theta \D \theta^2+\D \phi^2)
\quad {\rm with} \quad f_\pm(r_\pm)=&1-2M_\pm/r_\pm,
\label{ds}
\end{align} 
where quantities with the plus sign (the minus sign) belong to the outer
(inner) spacetime and $M_\pm$ are the masses in the inner and outer
regions, respectively.
We construct a spacetime having a single thin-shell at $\Sigma$,  on which the line element is given by
$\D s_{\Sigma}^2=h_{ab}\D y^a \D y^b:=-\D \tau^2+R(\tau)^2 (\sin^2\theta \D \theta^2+\D \phi^2)$,
where $\{y^a\}$ are the intrinsic coordinates on $\Sigma$ and are chosen as $y^i=(\tau, \theta, \phi)$.
$\tau$ stands for the proper time on the shell $\Sigma$ whose position is described by the coordinates $x^\mu(y^a)=(t(\tau),a(\tau),\theta, \phi)$. 
The equation of the surface $\Sigma$ is given by $r_\pm=R(\tau)$.  
The unit normals $n^\alpha$ to $\Sigma$ and basis vectors $e^\alpha_{a}:=\partial x^\alpha/\partial y^a$ tangent to $\Sigma$ are written by 
$n_{\alpha \pm}\D x^\alpha = -\dot{R}\D t+ \dot{t}_\pm \D r$, $u_\pm^\alpha \partial_{\alpha \pm}:= e_{\tau \pm}^\alpha \partial_{\alpha \pm}=  \dot{t}_\pm \partial_{t}+\dot{R} \partial_{r}$, $e^\alpha_{\theta}\partial_\alpha =\partial_{\theta}$, $e^\alpha_{\phi}\partial_\alpha =\partial_{\phi}$,
so that $u^{\alpha}u_{\alpha}=-1$, $n_\alpha n^\alpha=1$ and $u^\alpha
n_\alpha=0$ are satisfied, where the dot denotes the derivative with
respect to $\tau$.
The junction conditions are written as $[h_{ab}]=0$ and  
\begin{align}
8\pi S_{ab}=- [K_{ab}]+[K] h_{ab}, \label{hypersurface-eom}
\end{align}
where $K_{ab}:=h_{a}^{~c}h_{b}^{~d}\nabla_{c}n_{d}$ is the extrinsic
curvature, $S_{ab}$ is the stress-energy tensor confined on a
hypersurface $\Sigma$, and $[X]:=(X^+-X^-)|_{\Sigma}$.
The constraint equations are given by
\begin{align}
&S^{~b}_{a~|b}=- [T_{\alpha\beta}e^\alpha_{a}n^\beta], \label{hamiltonian-cons} \\
&\bar K^{ab}S_{ab}=  [T_{\alpha\beta}n^\alpha n^\beta], \label{momentum-cons}
\end{align}
where $\bar K^{ab}:=(K^{ab}_++K^{ab}_-)|_\Sigma/2$ and ``${}_{|a}$'' denotes
the covariant derivative associated with the induced metric $h_{ab}$ .
From $[h_{ab}]=0$, we obtain $\dot t_\pm:=F_\pm/f_\pm(R)$, where $F_\pm:=(f_\pm (R)+\dot{R}^2)^{1/2}$.
The non-zero components of the extrinsic curvature are 
$K_\tau^{\tau \pm}=\dot F_\pm/\dot{R}$ and $K_{\theta}^{\theta \pm}=K_{\phi}^{\phi \pm}=F_\pm/R$.
We take $S^i_j$ as a perfect fluid form, $S^i_j={\rm diag}(-\rho,p,p)$ with the surface pressure $p$ and the surface energy density $\rho$.
Then, \eq{hypersurface-eom} reduces to
\begin{align}
-4\pi \rho&=(F_+- F_-)/R, \label{JUNCTION1} \\
8\pi p&=(\dot F_+-\dot F_-)/\dot{R}+(F_+- F_-)/R. \label{JUNCTION2}
\end{align}
\eq{hamiltonian-cons} is explicitly  written by
\begin{align}
R\dot{\rho}=-2\dot{R}(p+\rho). \label{var-energy-cons}
\end{align}
From \eq{var-energy-cons}, $\rho$ is solved as $\rho=\rho(R)$ when the EOS is given. 
If the inner region is flat and the outer is Schwarzschild with a mass $M$, 
by reducing \eq{JUNCTION1}, we arrive $\dot R^2+V(R)=0$, where
\begin{align} 
V(R):=1-\frac{M}{R}-\left(\frac{M}{R}\right)^2\frac{1}{(4\pi \rho R)^2}-(2\pi \rho R)^2.  
\label{general-potential}
\end{align}
Here, we assumed the shell takes a linear EOS, i.e., $p=w\rho$ with $w=$const. 
By combining Eqs.~(\ref{JUNCTION1}) and (\ref{JUNCTION2}), we obtain
\begin{align}
F_+-F_-=C/R^{1+2w}, \label{JUNCTION3}
\end{align}
where $C$ is a constant of integration. Comparing \eq{JUNCTION1} and
\eq{JUNCTION3}, we can identify $C$ as
\begin{align}
-C=4\pi\rho R^{2(w+1)}=:m,
\end{align}
where $m$ is a constant. It should be noted that 
$4\pi R^{2}\rho ^{1/(w+1)}=(4\pi)^{w/(w+1)}m^{1/(w+1)}$ can be
identified with the conserved number 
of particles of which the shell consists.
Thus, the effective potential of \eq{general-potential} reduces to \eq{re-effective-potential}.

\end{document}